\documentclass[
prd,
longbibliography,
nofootinbib,
amsmath,
amssymb
]{revtex4-2}

\usepackage{mathtools}
\usepackage{bm}

\usepackage{graphicx}
\usepackage{xcolor}
\usepackage{booktabs}
\usepackage{mathrsfs}
\usepackage{array}

\usepackage{tikz}
\usetikzlibrary{arrows.meta,calc,positioning}

\usepackage[
colorlinks=true,
linkcolor=blue,
citecolor=blue,
urlcolor=blue
]{hyperref}

\usepackage[nameinlink,noabbrev]{cleveref}

\begin{document}

\title{Traces of Vacuum Squeezing: A Superconducting Transmission-Line
Analogue of Cosmological Particle Creation}

\author{Paula I. Villar}
\author{Federico Guglielmucci Nazar}
\author{Eli\'an Urtubey}
\author{Fernando C. Lombardo}
\affiliation{Departamento de Física ``Juan José Giambiagi'', Facultad de Ciencias Exactas y Naturales, Universidad de Buenos Aires, Buenos Aires, Argentina and Instituto de Física de Buenos Aires (IFIBA), CONICET--Universidad de Buenos Aires, Buenos Aires, Argentina}

\begin{abstract}
Particle creation in time-dependent quantum fields is a
central phenomenon in early-universe cosmology, yet it remains
experimentally inaccessible in its original context. We show theoretically that
superconducting circuits based on a waveguide-like transmission line
loaded with dc-SQUIDs can provide a direct analogue of a scalar field in
a radiation-dominated Friedmann--Robertson--Walker universe in
conformal time. The mode equation of the circuit reduces to the
Weber equation, the same equation governing cosmological particle
creation in a radiation-dominated background, admitting an exact
solution in terms of parabolic cylinder functions and an
exponentially suppressed particle spectrum. We analyse the conditions under which the analogy holds, which require
a compromise between the duration of the pulse, the uniformity of the
modulation and the symmetry of the SQUIDs, and show that dissipation
does not erase the cosmological signal. We  use the anomalous moment of each mode as a witness of the Bogoliubov mixing:
it vanishes for any thermal state and survives dissipation, and,
together with the photon number, it certifies squeezing, and with the purity of the state, it
identifies whether the pairs created from the vacuum dominate.
\end{abstract}

\maketitle

\section{Introduction}
\label{sec:intro}

One of the most striking predictions of quantum field theory in curved
spacetime is the creation of particles from the quantum vacuum driven
by a time-dependent gravitational
background~\cite{Parker1969,Birrell1982} (see Ref.~\cite{Ford2021}
for a modern review).
In a Friedmann--Robertson--Walker (FRW) universe the expansion of
spacetime mixes positive- and negative-frequency field modes, leading
to a non-trivial Bogoliubov transformation between asymptotic vacuum
states and a nonzero mean particle number in the out-region.
This mechanism amplifies primordial vacuum fluctuations into the seeds
of large-scale structure and plays a central role in inflationary
cosmology~\cite{Mukhanov1992}.
Yet the effect is extremely small in the present universe and, in the
early universe, permanently hidden from direct observation.

Laboratory analogues for curved-spacetime phenomena date back to
Unruh's sonic black-hole proposal~\cite{Unruh1981}.
In the cosmological context, analogues based on Bose--Einstein
condensates~\cite{Fedichev2003,Fischer2004,Barcelo2003,Horstmann2010},
ion traps~\cite{Alsing2005,Schutzhold2007,Wittemer2019},
and electromagnetic waveguides~\cite{Tian2017,Lang2019,Urtubey2026,Schutzhold2005,Nation2009}
have been proposed, and cosmological expansion has been simulated
experimentally in \mbox{Bose--Einstein} condensates~\cite{Eckel2018,Viermann2022}.
Further theoretical work on analogue particle creation in
curved-spacetime settings has been carried out in
Refs.~\cite{Lombardo2012,Lombardo2013}.
The dynamical Casimir effect (DCE), the generation of photon pairs
from the quantum vacuum by a time-dependent boundary condition,
provides a particularly clean circuit-QED implementation. It was  proposed
theoretically in a superconducting coplanar waveguide in Ref.~\cite{Johansson2009}, and later observed experimentally in Ref.~\cite{Wilson2011} in a coplanar waveguide terminated by
a flux-tunable SQUID. It was also reported 
in a Josephson metamaterial in Ref.~\cite{Lahteenmaki2013}, and an acoustic
analogue was observed in a Bose--Einstein condensate~\cite{Jaskula2012}; in circuits it can also entangle artificial atoms~\cite{Felicetti2014}.

Here, we propose an analogue of a specific cosmological model: a
spatially flat FRW universe dominated by radiation, which in conformal time
describes a contraction followed by an expansion
(Section~\ref{sec:cosmo}).
This choice has two distinguishing features.
First, the resulting mode equation is the Weber equation, which admits
an exact closed-form solution in terms of parabolic cylinder functions
and an analytically tractable exponentially suppressed particle
spectrum.
Second, the same Weber equation governs the Schwinger effect in
a constant electric field~\cite{Schwinger1951}, establishing a unified
mathematical framework connecting early-universe cosmology,
strong-field QED, and quantum optics within a single \mbox{experimental
platform.}

Following the waveguide architecture of Ref.~\cite{Tian2017}, we show
that a transmission line with a dc-SQUID in every cell reproduces, in
the continuum limit, the mode equation of the radiation-dominated
universe, with the Josephson term playing the role of the mass term of the
cosmological field.
The circuit and the inversion of the flux needed for a given profile
are those of Ref.~\cite{Tian2017}, which applied them to an asymptotically
static $\tanh$ expansion; the Weber equation is the
Landau--Zener problem, and its correspondence with the Schwinger effect
was worked out in Ref.~\cite{Rajeev2019}. What we propose is a different
analogue on the same platform: the radiation-dominated universe, whose
scale factor vanishes at $\eta=0$ and whose spectrum, fixed by the mass
alone, has no counterpart in the $\tanh$ model, together with its
open-system dynamics and an error budget that sets the hardware
tolerances.
Our contribution is threefold.
First, we  fix the parameters of the protocol for this specific
profile, so that the circuit inherits the Weber spectrum, which
depends only on the mass scale, and we determine the conditions under
which the analogy is faithful: a compromise between the duration of
the pulse, the uniformity of the modulation and the symmetry of the
SQUIDs.
Second, through a full treatment of photon loss, thermal excitation
and dephasing, we show that dissipation does not erase the
cosmological signal.
Third, building on the idea that field correlators are more robust
than the particle number~\cite{Lang2019}, we  use the anomalous moment $C_k$ of each standing-wave mode as a witness of the Bogoliubov mixing in the presence
of dissipation~\cite{Campo2005,Busch2014}, and establish, together with the photon number and the
purity of the state, when the mode is squeezed and when the pairs created from the vacuum dominate.

The paper is organized as follows.
Section~\ref{sec:cosmo} derives the Bogoliubov coefficients step by
step.
Section~\ref{sec:circuit} presents the circuit analogue, the
modulation protocol, and a numerical verification of the spectrum,
and discusses the connection with the Schwinger effect.
Section~\ref{sec:dissipation} derives the equations of motion
of the second-order moments from the Lindblad master equation and
analyses each dissipative channel.
Section~\ref{sec:correlations} introduces the anomalous moment and
the criteria for squeezing and for the dominance of the pairs created from the vacuum, and
derives their properties analytically.
Section~\ref{sec:conclusions} summarizes our results.
\section{Scalar Field in a Radiation-Dominated Universe}
\label{sec:cosmo}

We consider a spatially flat $(3+1)$-dimensional FRW spacetime,
\begin{equation}
  ds^{2} = -dt^{2} + a^{2}(t)\,d\bm{x}^{2},
\end{equation}
where $t$ is cosmic time and $a(t)$ is the scale factor.
In the radiation-dominated epoch one has $a(t)\propto t^{1/2}$.
Introducing conformal time $\eta$ via $d\eta = dt/a(t)$,  the metric becomes
\begin{equation}
  ds^{2} = a^{2}(\eta)\!\left(-d\eta^{2} + d\bm{x}^{2}\right),
  \label{eq:metric_conf}
\end{equation}
manifestly conformally flat. For radiation, $a(\eta)=a_1\eta$ and $ds^2=a_1^2\eta^2\left(-d\eta^2+d\bm{x}^2\right)$.
This conformal flatness has an important physical consequence: a
massless, conformally coupled scalar field in this background would
satisfy the same equation of motion as a free field in Minkowski
spacetime, and no particle creation would occur.
It is the mass term $m^2\phi^2$ that breaks conformal invariance and
generates a time-dependent effective mass in the mode equation,
giving rise to nonzero particle production.
Particle creation in the radiation-dominated universe is therefore a
purely massive effect, as we will see explicitly below.

We consider a real massive scalar field $\phi$ minimally coupled to
gravity, with action
\begin{equation}
  S = -\frac{1}{2}\int d^{4}x\,\sqrt{-g}\,
      \left(g^{\mu\nu}\partial_\mu\phi\,\partial_\nu\phi
      + m^{2}\phi^{2}\right).
\end{equation}
The equation of motion $\Box\phi - m^2\phi = 0$ in conformal
coordinates reads
$\phi'' + 2(a'/a)\phi'$ $- \nabla^2\phi + m^2 a^2 \phi = 0$,
where primes denote $\partial_\eta$.
Introducing the rescaled field $\chi = a\phi$ eliminates the
first-derivative term.
For $a\propto\eta$ one has $a'' = 0$ exactly (and hence
\mbox{$R=6a''/a^3=0$}, so that minimal and conformal coupling coincide). Absorbing $a_1$ into $m$, the spatial Fourier
amplitudes $\chi_k(\eta)$ satisfy
\begin{equation}
  \chi_k'' + \omega_k^{2}(\eta)\,\chi_k = 0,
  \qquad
  \omega_k^{2}(\eta) = k^{2} + m^{2}\eta^{2}.
  \label{eq:mode}
\end{equation}
This is a harmonic oscillator with a time-dependent frequency that
reaches its minimum at $\eta=0$, the non-adiabatic region where
particle creation is concentrated. Equation~\eqref{eq:mode} depends on the scale factor only
through $a^2=a_1^2\eta^2$, and we consider it on the whole
conformal-time axis, $\eta\in(-\infty,\infty)$, with the in-vacuum
defined at $\eta\to-\infty$.
The background is then a radiation-dominated contraction ($\eta<0$,
$a\propto(-t)^{1/2}$) followed by a radiation-dominated expansion
($\eta>0$, $a\propto t^{1/2}$): the two branches of the Friedmann
equation for radiation, joined at $a=0$.
The rescaled field $\chi=a\phi$ remains regular at $\eta=0$, since only
the analytic function $a^2$ enters Equation~\eqref{eq:mode}, and the
evolution through the bounce is well defined.

The substitution $z = \sqrt{2m}\,e^{i\pi/4}\eta$ (so that
$z^2 = 2im\eta^2$) transforms Equation~\eqref{eq:mode} into the standard
Weber (parabolic cylinder) equation,
\begin{equation}
  \frac{d^{2}\chi_k}{dz^{2}} +
  \left(\nu + \frac{1}{2} - \frac{z^{2}}{4}\right)\chi_k = 0,
  \qquad
  \nu = -\frac{1}{2} - i\lambda,\quad
  \lambda = \frac{k^{2}}{2m},
  \label{eq:weber}
\end{equation}
whose solutions are the parabolic cylinder functions $D_\nu(z)$.
The parameter $\lambda$ measures the ratio of the squared wavevector
to the mass scale and controls the exponential suppression of particle
production, as we will see below.

The field $\chi$ is promoted to an operator and expanded in mode
functions $u_k(\eta)$,
\begin{equation}
  \hat\chi(\eta,\bm{x}) = \int\!\frac{d^3k}{(2\pi)^3}
  \bigl[\hat{a}_{\bm{k}}\,u_k(\eta)\,e^{i\bm{k}\cdot\bm{x}}
       +\hat{a}^\dagger_{\bm{k}}\,u_k^*(\eta)\,e^{-i\bm{k}\cdot\bm{x}}\bigr],
\end{equation}
subject to the Wronskian normalization
$u_k u_k^{*\prime} - u_k^* u_k' = i$.
The in-vacuum $|{\rm in}\rangle$ is defined by
$\hat{a}_{\bm k}|{\rm in}\rangle=0$, corresponding to the absence of
particles as seen by inertial observers as $\eta\to-\infty$.
The positive-frequency out-modes, satisfying the WKB condition
\mbox{$u_k^{\rm out}\sim e^{-i\int\omega_k d\eta}/\sqrt{2\omega_k}$} as
$\eta\to+\infty$, are given exactly by

\begin{equation}
  u_k^{{\rm out}}(\eta) = \frac{e^{-\pi\lambda/4}}{(2m)^{1/4}}\,D_\nu(z).
  \label{eq:inmodes}
\end{equation}
In the asymptotic past, in-observers define modes
$u_k^{\rm in}$ via the WKB condition as $\eta\to-\infty$.
Because the FRW background is spatially homogeneous, modes with
different wavevectors $\bm{k}$ and $\bm{k}'$ evolve independently:
the Bogoliubov transformation is diagonal in $\bm{k}$, with each
mode $k$ mixing only with $-k$ (by momentum conservation).
This is the \mbox{Bogoliubov transformation}
\begin{equation}
  u_k^{\rm in} = \alpha_k\,u_k^{\rm out} + \beta_k\,u_k^{{\rm out}*},
  \label{eq:bogo}
\end{equation}
with $|\alpha_k|^2-|\beta_k|^2=1$.
Using the connection formula for $D_\nu$,
\begin{equation}
  D_\nu(z) = e^{-i\pi\nu}D_\nu(-z)
  + \frac{\sqrt{2\pi}}{\Gamma(-\nu)}\,e^{-i\pi(\nu+1)/2}D_{-\nu-1}(iz),
\end{equation}
and the asymptotics $D_\nu(z)\sim z^\nu e^{-z^2/4}$, one identifies 
as \mbox{$\eta\to-\infty$, $D_{-\nu-1}(iz)\sim e^{+im\eta^2/2}$} as the in
positive-frequency mode, and $D_\nu(-z)\sim e^{-im\eta^2/2}$ as the in
negative-frequency mode. The connection formula thus expresses the
out-mode in terms of the in-modes, and inverting it gives $\alpha_k$
and $\beta_k$.
With $\Gamma(-\nu)=\Gamma(1/2+i\lambda)$ and the identity
\mbox{$|\Gamma(1/2+i\lambda)|^2=\pi/\cosh(\pi\lambda)$}, one
obtains~\cite{Birrell1982}
\begin{equation}
  N_k \equiv |\beta_k|^{2} = e^{-\pi A},
  \qquad A \equiv \frac{k^{2}}{m} = 2\lambda,
  \label{eq:Nk_cosmo}
\end{equation}
where $A$ is the dimensionless Weber parameter controlling the
exponential suppression of the spectrum.
The mean particle number $N_k = |\beta_k|^2$ follows from the
Bogoliubov structure: since $\hat a_k|{\rm in}\rangle=0$, the
expected number of out-particles is
$\langle{\rm in}|\hat{b}^\dagger_k\hat{b}_k|{\rm in}\rangle=|\beta_k|^2$,
where $\hat b_k=\alpha_k\hat a_k+\beta_k^*\hat a_{-k}^\dagger$
are the out-annihilators.
Nonzero $\beta_k$ signals that the in-vacuum contains out-particles:
the time-dependent background has mixed creation and annihilation
operators, populating the out-state with pairs.

This confirms the statement made above: for $m\to0$,
$A=k^2/m\to\infty$ for any fixed $k\neq0$ and $N_k\to0$, since
$\omega_k^2=k^2$ is then constant and the in- and out-vacua coincide.
The mass term $m^2\eta^2$ in Equation~\eqref{eq:mode} is what makes the
evolution non-adiabatic near $\eta=0$.

The spectrum of Equation~\eqref{eq:Nk_cosmo}, is a Gaussian in $k$, exponentially
suppressed for \mbox{$k^2\gg m$ \cite{Birrell1982,Parker1969}}.
Significant production occurs only for $k^2\lesssim m$, where the
adiabaticity condition $|\omega_k'/\omega_k^2|\ll1$ breaks down near
$\eta=0$.
The suppression has a semiclassical interpretation:
$N_k\sim e^{-2\,\mathrm{Im}\,S_{\rm WKB}}$ where the imaginary WKB
action between the complex turning points of $\omega_k^2(\eta)=0$
gives $\mathrm{Im}\,S_{\rm WKB}=\pi k^2/(2m)$ \cite{Birrell1982}.

This contrasts with the widely studied tanh model, where
\mbox{$a^2(\eta)=A_0+B\tanh(\rho\eta)$ \cite{Birrell1982,Tian2017}.}
In that case the particle spectrum is
\begin{equation}
  N_k^{\rm tanh} = \frac{\sinh^2(\pi\omega_-/\rho)}
                        {\sinh(\pi\omega_{\rm in}/\rho)\,\sinh(\pi\omega_{\rm out}/\rho)},
  \label{eq:Nk_tanh}
\end{equation}
with $\omega_\pm=(\omega_{\rm out}\pm\omega_{\rm in})/2$, which has a
free timescale parameter $\rho$ and approaches a thermal-like
distribution $N_k\approx e^{-2\pi\omega_{\rm in}/\rho}$, in the
slow limit $\omega_{\rm in},\omega_{\rm out}\gg\rho$.
By contrast, the radiation-dominated spectrum of
Equation~\eqref{eq:Nk_cosmo} has no free timescale: the only scale is $m$.
This makes the cosmological spectrum particularly predictive and clean
for \mbox{analogue purposes.}

\section{Radiation-Dominated Universe: Transmission Line Analogue}
\label{sec:circuit}

\begin{figure}[h!]
  \begin{minipage}{0.56\textwidth}
  \centering
  \includegraphics[width=\linewidth]{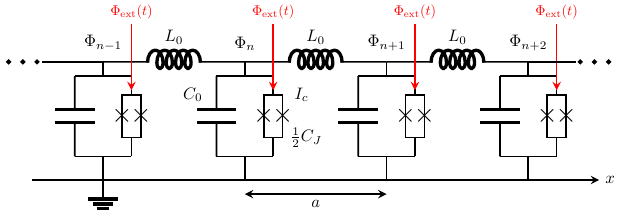}
  \\[4pt] (a)
  \end{minipage}%
  \hfill
  \begin{minipage}{0.38\textwidth}
  \centering
  \includegraphics[width=\linewidth]{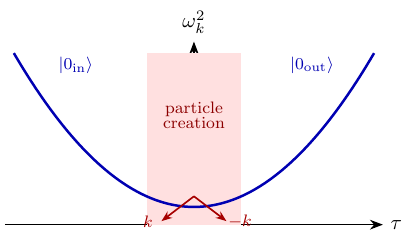}
  \\[4pt] (b)
  \end{minipage}
  \caption{%
    (\textbf{a}) Circuit of
 the analogue. Each cell of the transmission line
    (length $a$) has a series inductance $L_0$ and a shunt formed by a
    capacitor $C_0$ in parallel with a dc-SQUID ($\times$: Josephson
    junctions). The external flux $\Phi_{\rm ext}(t)$ (red arrows)
    tunes the Josephson inductance $L_J(t)$ of every SQUID.
    (\textbf{b}) Mode frequency $\omega_k$ as a function of time, in cosmology
    (conformal time $\eta$) and in the circuit (laboratory time $t$).
    Far from $\tau=0$ the frequency varies slowly and the in and out
    vacua are well defined; inside the shaded band the evolution is
    non-adiabatic and creates pairs $(k,-k)$ (\mbox{red arrows}).
  }
  \label{fig:circuit}
\end{figure}

We adopt the coplanar transmission line architecture of
Ref.~\cite{Tian2017} (Figure~\ref{fig:circuit}), in which every
shunt capacitor $C_0$ is paralleled by an identical dc-SQUID with
critical current $I_c$, one per unit cell of length $a$, rather than
a single SQUID placed at one end as in conventional DCE experiments
(Ref.~\cite{Wilson2011}).
This distributed modulation acts identically on every
cell: a single terminating SQUID
would modulate only the boundary condition, coupling modes of
different wavevectors and producing a non-diagonal Bogoliubov
transformation, whereas uniformly distributed SQUIDs modulate the
bulk homogeneously, preserving the independence of each mode, exactly
as required to reproduce the homogeneous FRW background.
Unlike Ref.~\cite{Tian2017}, which considered a generic
$\tanh$-shaped scale factor with a free rapidity parameter $\rho$ (and gave the
flux inversion used below, its Equation~(11)), in order to reproduce the radiation-dominated profile $a^2(\eta)\propto\eta^2$, whose spectrum depends only on the mass scale, the circuit adds a single scale, the duration of the pulse, which must be long enough for the Weber result to hold (see Appendix~\ref{app:validity}).

The node flux $\Phi_n$ at site $n$ is the dynamical variable;
the goal is to show that in the continuum limit the mode amplitudes
satisfy an equation of the same Weber form as Equation~\eqref{eq:mode},
from which the particle spectrum follows once the flux protocol is
fixed.

The starting point is the discrete Lagrangian of the circuit.
For a chain of $N$ identical cells, it reads
\begin{equation}
  \mathscr{L} = \sum_{n=1}^{N}\!\left[
    \frac{C}{2}\dot\Phi_n^2
    - \frac{(\Phi_n-\Phi_{n-1})^2}{2L_0}
    + E_J(\Phi_{\rm ext})\cos\!\left(\frac{2\pi\Phi_n}{\Phi_0}\right)
  \right],
  \label{eq:lagrangian_discrete}
\end{equation}
where $C=C_0+C_J$ is the total shunt capacitance per cell (shunt
capacitor plus junction capacitance) and $\Phi_n$ is the node flux at site $n$.
The first term is the kinetic (electric) energy stored in $C$; the
second is the inductive (magnetic) energy of $L_0$; and the third is
the Josephson energy of each dc-SQUID.

The dc-SQUID consists of two identical Josephson junctions in a
superconducting loop.
When the external flux $\Phi_{\rm ext}(t)$ threads the loop, the
two junctions combine into a single effective junction with tunable
Josephson energy
$E_J(\Phi_{\rm ext})=E_J^{\max}|\cos(\pi\Phi_{\rm ext}/\Phi_0)|$,
where $E_J^{\max}=\Phi_0 I_c/(2\pi)$, $I_c$ is the critical current of
the dc-SQUID (the sum of the two junction critical currents), and
$\Phi_0=h/2e$ is the flux quantum.
We work in the phase regime where the node-flux oscillations are
small, $2\pi\Phi_n/\Phi_0\ll1$ (Appendix~\ref{app:validity}), so the Josephson cosine expands to
quadratic order in $\Phi_n$:

\begin{equation}
  E_J(\Phi_{\rm ext})\cos\!\left(\frac{2\pi\Phi_n}{\Phi_0}\right)
  \approx E_J(\Phi_{\rm ext}) - \frac{E_J(\Phi_{\rm ext})}{2}
  \left(\frac{2\pi\Phi_n}{\Phi_0}\right)^{\!2}
  = \mathrm{const} - \frac{\Phi_n^2}{2L_J(t)},
\end{equation}
where the first term does not depend on $\Phi_n$ and drops out of
the equations of motion, and the last step defines the tunable Josephson inductance
\begin{equation}
  L_J(t) \equiv \frac{\Phi_0^2}{4\pi^2 E_J(\Phi_{\rm ext}(t))}
         = \frac{\Phi_0}{2\pi I_c\,|\cos(\pi\Phi_{\rm ext}(t)/\Phi_0)|}.
  \label{eq:LJ}
\end{equation}
This inductance diverges when $\Phi_{\rm ext}=\Phi_0/2$ (where the two junctions interfere destructively) and decreases
as the flux departs from this value, providing the tunable effective
mass for the analogue.

Taking the continuum (long-wavelength) limit $ka\ll1$, with
$D=Na$, $\Phi_n(t)\to\Phi(x,t)$ and
$\Phi_n-\Phi_{n-1}\to a\,\partial_x\Phi$, the sum becomes an
integral and the Lagrangian reduces to
\begin{equation}
  \mathscr{L} = \frac{C}{2a}\int_0^D\! dx\left[
    \!\left(\frac{\partial\Phi}{\partial t}\right)^{\!2}
    - \frac{a^2}{L_0 C}\left(\frac{\partial\Phi}{\partial x}\right)^{\!2}
    - \frac{1}{L_J(t)C}\,\Phi^2
  \right],
  \label{eq:lagrangian}
\end{equation}
with propagation speed $v=a/\sqrt{L_0C}$.
The Euler--Lagrange equation for $\Phi(x,t)$ is
\begin{equation}
  \frac{\partial^2\Phi}{\partial t^2}
  - v^2\frac{\partial^2\Phi}{\partial x^2}
  + \frac{1}{L_J(t)C}\,\Phi = 0.
  \label{eq:eom}
\end{equation}
This is a wave equation with a spatially uniform, time-dependent
potential term $1/[L_J(t)C]$ controlled by the external flux.
We expand the field in spatial normal modes,
\mbox{$\Phi(x,t)=\sum_k q_k(t)\,u_k(x)$,}
where the $u_k(x)$ satisfy the spatial eigenvalue problem
\begin{equation}
  -v^2\frac{d^2 u_k}{dx^2} = \omega_{k,{\rm kin}}^2\,u_k,
  \label{eq:spatial_eig}
\end{equation}
with Neumann (open) boundary conditions at $x=0$ and $x=D$,
$\partial_x u_k\big|_{x=0,D}=0$.
These correspond to open ends of the transmission line, appropriate
for a line that is not terminated by any fixed element.
The normalised solutions are $u_k(x)=\sqrt{2/D}\cos(kx)$ (with
$u_0=1/\sqrt{D}$ for $k=0$) and the allowed wavevectors are
$k=n\pi/D$ for non-negative integer $n$.
The eigenfunctions satisfy the orthogonality relation
$\int_0^D u_k(x)u_{k'}(x)\,dx = \delta_{kk'}$.

Substituting the expansion into Equation~\eqref{eq:eom}, multiplying by
$u_{k'}(x)$, and integrating over $[0,D]$, the spatial kinetic
term $-v^2\partial_{xx}\Phi$ projects onto $v^2k'^2 q_{k'}$ by
virtue of \mbox{Equation~\eqref{eq:spatial_eig}}.
The crucial observation is that the coefficient $[L_J(t)C]^{-1}$ of the Josephson potential term
 is spatially uniform: since $L_J(t)$ is the
same for every cell, this term projects as
$[L_J(t)C]^{-1}\int_0^D u_{k'}(x)\Phi\,dx
=[L_J(t)C]^{-1}q_{k'}$,
without coupling modes of different wavevectors.
This is the mathematical expression of the physical argument made
above: a spatially uniform modulation cannot mix modes.
The result is that each amplitude $q_k(t)$ evolves independently,
satisfying
\begin{equation}
  \ddot{q}_k + \omega_k^{2}(t)\,q_k = 0,
  \qquad
  \omega_k^{2}(t) = v^2 k^{2} + \frac{1}{L_J(t)C}.
  \label{eq:circuit_mode}
\end{equation}
This is a harmonic oscillator equation with a time-dependent
frequency: the first term $v^2k^2$ is the dispersion of the bare
transmission line, and the second term $1/[L_J(t)C]$ is the
time-dependent contribution from the Josephson inductance
(Figure~\ref{fig:circuit}), which plays the role of an effective mass.

Comparing Equations~\eqref{eq:mode} and~\eqref{eq:circuit_mode}, the
two equations are structurally identical under the dictionary of
Table~\ref{tab:dictionary}.

\begin{table}[t]
  \caption{Correspondence between the cosmological and circuit
  descriptions: the two mode equations,
  Equations~\eqref{eq:mode} and~\eqref{eq:circuit_mode}, are identical
  under this correspondence.}
  \begin{tabular}{ll}
    \toprule
    \textbf{Cosmology} & \textbf{Circuit} \\
    \midrule
    $\eta$ (conformal time) & $t$ (laboratory time) \\
    $k^2$ & $v^2 k^2$ \\
    $m$ & $m_{\rm eff}$ \\
    $m^2\eta^2$ & $1/[L_J(t)C]$ \\
    $A=k^2/m$ (Weber parameter) & $\omega_{k0}^2/m_{\rm eff}$ \\
    \bottomrule
  \end{tabular}
  \label{tab:dictionary}
\end{table}

Laboratory time $t$ plays the role of conformal time $\eta$, not cosmic time: in conformal time, and because $a''=0$ for $a\propto\eta$ (Section \ref{sec:cosmo}), the rescaled field obeys a harmonic-oscillator equation without friction term, which is the equation the circuit reproduces (within the conditions discussed in Appendix~\ref{app:weber}).

For the analogy to reproduce Equation~\eqref{eq:Nk_cosmo}, one requires
$1/[L_J(t)C]=m_{\rm eff}^2 t^2$.
From Equation~\eqref{eq:LJ}, this is achieved by the flux protocol
\begin{equation}
  \Phi_{\rm ext}(t) =
  \frac{\Phi_0}{\pi}\arccos\!\left(\frac{t^2}{t_*^2}\right),
  \quad|t|\leq t_*,
  \label{eq:protocol}
\end{equation}
where $t_*$ is the modulation window.
At $t=0$, $\Phi_{\rm ext}=\Phi_0/2$ ($L_J\to\infty$, Josephson term off); as $|t|$ increases, $L_J(t)$
decreases and the Josephson term grows quadratically.
At $|t|=t_*$ the flux reaches zero and the Josephson
term its maximum, $\omega_{\max}^2=1/(L_{J,\min}C)=m_{\rm eff}^2t_*^2$;
outside the window the flux is held at zero and the system evolves
adiabatically.
The effective mass is
\begin{equation}
  m_{\rm eff} = \sqrt{\frac{2\pi I_c}{\Phi_0\,C\,t_*^2}}.
  \label{eq:meff}
\end{equation}
The same Weber parameter $A$ of Equation~\eqref{eq:Nk_cosmo} now reads
$A = \omega_{k0}^{2}/m_{\rm eff}$ with $\omega_{k0}=vk$, under the
correspondence of Table~\ref{tab:dictionary}, giving the circuit particle
spectrum in exact formal correspondence with Equation~\eqref{eq:Nk_cosmo},
with the circuit's mass $m_{\rm eff}$ vanishing when $t_*\to\infty$
(no modulation).
By varying $m_{\rm eff}$ (changing $t_*$ and $I_c$ together so
that $\omega_{\max}^2/m_{\rm eff}$ stays fixed) while fixing
$\omega_{k0}$, one sweeps $A$ continuously and maps out the
spectrum (see Appendix~\ref{app:weber}).

\begin{figure}[h!]
  \includegraphics[width=0.7\textwidth]{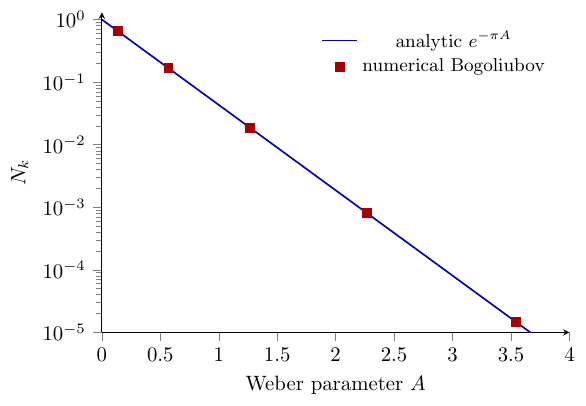}
  \caption{%
    Mean photon number $N_k=|\beta_k|^{2}$ vs. the Weber parameter
    $A$ for the ideal profile of \mbox{Equation~\eqref{eq:mode_dimless}.}
    Blue: analytic prediction $e^{-\pi A}$;
    red squares: numerical Bogoliubov extraction, integrating
    Equation~\eqref{eq:mode_dimless} directly and projecting onto the
    WKB out-mode.
    The agreement is within $1\%$ in all cases, down to
    $N_k\sim10^{-5}$.
  }
  \label{fig:spectrum}
\end{figure}

In cosmological language, the protocol of Equation~\eqref{eq:protocol}
realises a radiation-dominated contraction and re-expansion
of finite duration, through $a=0$ at $t=0$ ($\Phi_{\rm ext}=\Phi_0/2$),
embedded between two static eras in which the in and out vacua are well
defined~\cite{Birrell1982}. For asymmetric SQUIDs the Josephson term does not vanish
at $\Phi_0/2$, and the circuit realises instead a regular bounce, with
$a_{\min}^2\propto d$. 
In dimensionless time $\tau=\sqrt{m_{\rm eff}}\,t$, the mode equation
reads
\begin{equation}
  \frac{d^{2}q_k}{d\tau^{2}} + (A + \tau^{2})\,q_k = 0.
  \label{eq:mode_dimless}
\end{equation}
We integrate numerically from WKB initial conditions at
$\tau_i\ll-\sqrt{A}$ to $\tau_f\gg\sqrt{A}$ and extract $\beta_k$
from the Wronskian overlap with the out-mode,
\begin{equation}
  \beta_k = i\bigl[u(\tau_f)\,v'(\tau_f) - u'(\tau_f)\,v(\tau_f)\bigr],
\end{equation}
where $v(\tau)$ is the WKB out-mode with accumulated phase
$\Phi=\int_{\tau_i}^{\tau_f}\!\sqrt{A+\tau'^2}\,d\tau'$.
Figure~\ref{fig:spectrum} shows the results for five
illustrative values of the Weber parameter, for the
ideal profile of Equation~\eqref{eq:mode_dimless} without finite-window
corrections.
Numerical extraction agrees with $e^{-\pi A}$ to within $1\%$ in
all cases, confirming the exponential dependence on $A$
(Gaussian in $k$) down to $N_k\sim10^{-5}$.

\subsection*{Connection with the Schwinger Effect}

Beyond the cosmological analogue, the same Weber equation,
Equation~\eqref{eq:mode}, governs a well-known non-perturbative
phenomenon in quantum electrodynamics: pair production by a
constant electric field, the Schwinger effect
(Ref.~\cite{Schwinger1951}).
This is not a coincidence: in Ref.~\cite{Rajeev2019} the authors
established a general algebraic correspondence between cosmological
particle production and the Schwinger effect, of which the shared
Weber equation here is a specific realisation.
 We recall it here because, with its
parameters reinterpreted, the same circuit gives the Schwinger spectrum
for a single momentum sector. Related gravitational analogues to the
Schwinger effect in conformally-flat spacetimes have also been
studied in Ref.~\cite{FranchinoVinas2026}.
Consider a charged scalar field $\phi$ of mass $m$ and charge $e$
in a constant, linearly polarised electric field $\bm{E}=E\hat x$.
The field acts on the charged scalar by shifting its kinetic momentum:
in the temporal gauge with vector potential $A_x(t)=-Et$, the
Klein--Gordon equation for the Fourier modes reads
\begin{equation}
  \ddot\phi_{\bm{k}} +
  \bigl[m^2 + k_\perp^2 + (k_\parallel - eEt)^2\bigr]\phi_{\bm{k}} = 0,
  \label{eq:schwinger}
\end{equation}
where $k_\parallel$ ($k_\perp$) is the momentum component parallel
(perpendicular) to $\bm{E}$.
What is created are pairs of charged scalar bosons (particle and
antiparticle) from the quantum vacuum.
The constant electric field provides the energy needed to materialise
these virtual pairs: modes whose canonical momentum $k_\parallel-eEt$
passes through zero gain sufficient energy from the field to become
real on-shell particles.
Defining $m_\perp^2=m^2+k_\perp^2$ and shifting
the time origin to $t_0=k_\parallel/(eE)$ (the moment of minimum
kinetic energy for that mode), Equation~\eqref{eq:schwinger} reduces to
\begin{equation}
  \ddot\phi_{k_\parallel} +
  \bigl[m_\perp^2 + (eE)^2(t-t_0)^2\bigr]\phi_{k_\parallel} = 0,
\end{equation}
which is the Weber equation~\eqref{eq:mode_dimless} with
{$A=m_\perp^2/(eE)$}.
The Bogoliubov coefficient for pair production is
{$|\beta|^2=e^{-\pi m_\perp^2/(eE)}$}, in formal analogy with
Equation~\eqref{eq:Nk_cosmo}.
The physical picture is identical in both cases: a mode frequency
that passes through a minimum at finite time ($t=t_0$ in Schwinger,
$\eta=0$ in cosmology), near which the adiabatic approximation breaks
down and particle pairs are created from the vacuum.

Regarding the polarisation: for the homogeneous field considered here
the spectrum is independent of the orientation of $\bm{E}$.
In $(3+1)$ dimensions, integrating over transverse momenta $k_\perp$
gives the full Schwinger rate $\propto (eE)^2 e^{-\pi m^2/(eE)}$.
The circuit analogue reproduces the mode dynamics for a single
$(k_\parallel,k_\perp=0)$ sector; it does not simulate the full
$(3+1)$-dimensional integral over $k_\perp$, and consequently the
prefactor of the Schwinger rate is not reproduced.
The exponential structure, which encodes the essential
non-perturbative physics, is faithfully captured.

\section{Dissipative Radiation-Dominated Transmission Line}
\label{sec:dissipation}

In this section we compute the particle spectrum of the transmission
line for the circuit parameters of Table~\ref{tab:params}
(Appendix~\ref{app:validity}): mode $n=1$,
$\omega_{k0}/2\pi\approx0.35\,\mathrm{GHz}$, $A\approx0.14$ and
$t_*\approx0.85\,\mathrm{ns}$; unless stated otherwise, the SQUIDs are
taken to be symmetric.
Any implementation of the analogue is, however, coupled to its
electromagnetic environment.
For completeness, we analyse how the three main sources of dissipation
and noise (photon loss, thermal excitation and dephasing) modify the
created spectrum.
The standard framework for open quantum systems in the Markovian
regime is the Lindblad master equation (Ref.~\cite{Breuer2002}), which
we use as an effective description of the model. It mainly
assumes weak system--bath coupling and a memoryless bath
(see Appendix~\ref{app:lindblad}).
For a uniform modulation (Appendix~\ref{app:mixing}) the modes evolve
independently, and each couples to its own reservoir, so the problem
reduces to a family of single-mode problems.

\begin{figure}[h!]
  \begin{minipage}[b]{0.316\textwidth}\centering
  \includegraphics[width=\linewidth]{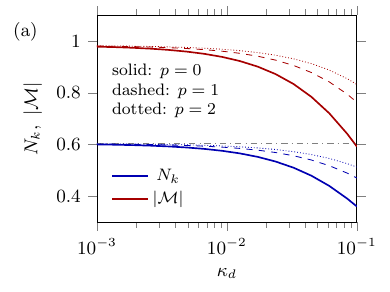}\end{minipage}\hfill
  \begin{minipage}[b]{0.316\textwidth}\centering
  \includegraphics[width=\linewidth]{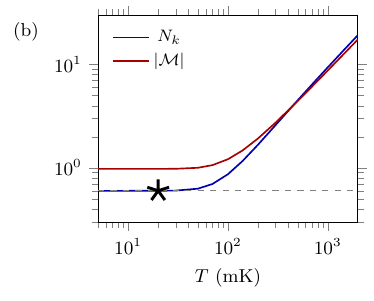}\end{minipage}\hfill
  \begin{minipage}[b]{0.316\textwidth}\centering
  \includegraphics[width=\linewidth]{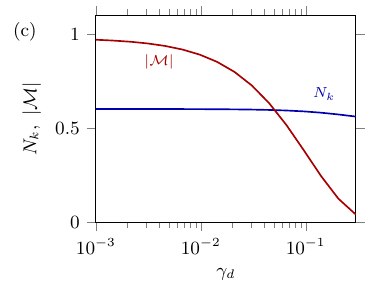}\end{minipage}
  \caption{%
    Effect
 of dissipation on the signal for mode $n=1$, for the
    parameters of Table~\ref{tab:params}.
    (\textbf{a}) $N_k$ (blue) and $|\mathcal{M}|$ (red) vs. dimensionless loss
    $\kappa_d$ at $n_{\rm th}=0$; the dashed lines (panels a
    and b) mark the unitary value of the protocol, $N_k\approx0.60$.
    Both stay within a few percent of this value for
    $\kappa_d\leq0.01$.
    Dashed and dotted: $\kappa\propto\omega_k^p$ with $p=1$ and $2$,
    with $\kappa$ fixed at the readout frequency.
    (\textbf{b}) $N_k$ and $|\mathcal{M}|$ vs. physical temperature at fixed
    $\kappa_d=0.001$, starting from thermal equilibrium, with
    $n_{\rm th}$ evaluated at the instantaneous frequency; the star
    marks the operating point, $T=20\,\mathrm{mK}$, a typical base
    temperature of the dilution refrigerators used in circuit-QED
    experiments (Table~\ref{tab:params}).
    Thermal photons dominate $N_k$ only above
    $\approx145\,\mathrm{mK}$.
    (\textbf{c}) $N_k$ and $|\mathcal{M}|$ vs. dimensionless dephasing rate
    $\gamma_d$ at the operating point.
  }
  \label{fig:dissipation}
\end{figure}

The three physical dissipation channels are the following.
Photon loss at rate $\kappa$ has jump operator
$L_1=\sqrt{\kappa(n_{\rm th}+1)}\,\hat{a}$ and models energy leakage
into the electromagnetic environment via the input/output ports or
residual resistive losses.
We take $\kappa$ independent of the instantaneous frequency,
which corresponds to an ohmic environment, such as a resistive load or
a semi-infinite transmission line.
For a capacitively coupled port $\kappa\propto\omega_k^p$ with $p=1$ or
$2$; with $\kappa$ fixed at the readout frequency (Table~\ref{tab:params}),
this only lowers the loss during the crossing, so a constant $\kappa$ is
the conservative case (Figure~\ref{fig:dissipation}a).
Thermal excitation at rate $\kappa n_{\rm th}$ has jump operator
$L_2=\sqrt{\kappa n_{\rm th}}\,\hat{a}^\dagger$, where
$n_{\rm th}=1/(e^{\hbar\omega_k{(t)}/k_BT}-1)$ is the mean thermal
occupation of the bath at temperature $T$, evaluated at the
instantaneous mode frequency $\omega_k(t)$.
These two channels together drive the mode toward thermal equilibrium
with mean occupation $n_{\rm th}$.
Dephasing at rate $\gamma_\phi$ has jump operator
$L_3=\sqrt{\gamma_\phi}\,\hat{a}^\dagger\hat{a}$ and models flux
noise in the SQUID loops, which randomly shifts the mode frequency
and destroys phase coherence without changing the photon number.
With these considerations, the Lindblad master equation
for mode $k$ reads
\begin{align}
  \frac{d\rho_k}{dt} &= -i[H_k(t),\rho_k]
  + \kappa(n_{\rm th}+1)\mathcal{D}[\hat{a}_k]\rho_k
  + \kappa n_{\rm th}\,\mathcal{D}[\hat{a}_k^\dagger]\rho_k
  + \gamma_\phi\,\mathcal{D}[\hat{a}_k^\dagger\hat{a}_k]\rho_k,
  \label{eq:lindblad}
\end{align}
where $H_k(t)=\omega_k(t)(\hat a_k^\dagger\hat a_k+1/2)$ is the
instantaneous Hamiltonian of mode $k$ and
\mbox{$\mathcal{D}[\hat{O}]\rho=\hat{O}\rho\hat{O}^\dagger
-\tfrac{1}{2}\hat{O}^\dagger\hat{O}\rho
-\tfrac{1}{2}\rho\hat{O}^\dagger\hat{O}$.}

The initial state is Gaussian (the vacuum, or a thermal
state when $n_{\rm th}>0$), and loss and thermal excitation preserve
this character; dephasing does not, but the equations for the
second-order moments below remain closed.
Since there is no coherent drive, $\langle\hat a(t)\rangle=0$, and the
quantities of interest are the two second-order moments of mode $k$ (we drop the index $k$ in what follows)
\begin{equation}
  n(t) \equiv \langle\hat{a}^\dagger(t)\hat{a}(t)\rangle,
  \qquad
  \mathcal{M}(t) \equiv \langle\hat{a}(t)\hat{a}(t)\rangle,
  \label{eq:moments}
\end{equation}
where $\hat{a}(t)$ is the annihilation operator of the instantaneous
oscillator of frequency $\omega_k(t)$, that is, the one that
diagonalises $H_k(t)$; in the circuit, its quadratures are
proportional to the flux amplitude $q_k$ of mode $k$ and to its
conjugate momentum.

The quantity $n(t)$ is the mean photon number and $\mathcal{M}(t)$
is the anomalous moment, measuring the degree of two-photon coherence
(squeezing) in the mode.

To derive equations for $n$ and $\mathcal{M}$, we differentiate
Equation~\eqref{eq:moments} and use Equation~\eqref{eq:lindblad}.
Since $\omega_k(t)$ varies, the unitary part of the evolution of the instantaneous ladder operators mixes
with their conjugates,
\begin{equation}
  \frac{d\hat{a}}{dt}\bigg|_{\rm cons.}
  = -i\omega_k(t)\hat{a} + \xi(t)\hat{a}^\dagger,
  \qquad
  \xi(t) = \frac{\dot\omega_k(t)}{2\omega_k(t)},
  \label{eq:da_dt}
\end{equation}
where $\xi(t)$ is the non-adiabaticity parameter.
The term $\xi(t)\hat{a}^\dagger$ is the source of squeezing and
particle creation even in the absence of dissipation: a
time-varying frequency rotates the instantaneous vacuum, generating
two-photon coherences.
Combining Equation~\eqref{eq:da_dt} with the dissipators of
Equation~\eqref{eq:lindblad}, and working in dimensionless time
$\tau=\sqrt{m_{\rm eff}}\,t$ with
$\Omega(\tau)=\sqrt{A+\tau^2}$ for
$|\tau|\leq\tau_*=\sqrt{m_{\rm eff}}\,t_*$ and constant outside,
$\kappa_d=\kappa/\sqrt{m_{\rm eff}}$,
$\gamma_d=\gamma_\phi/\sqrt{m_{\rm eff}}$, and
$\xi(\tau)=\tau/[2(A+\tau^2)]$ inside the window and zero
outside, one arrives at the moment equations
\begin{align}
  \dot{n} &= -\kappa_d\,n
             + \kappa_d\,n_{\rm th}(\tau)
             + 2\xi(\tau)\,\mathrm{Re}(\mathcal{M}),
  \label{eq:bl_n}\\
  \dot{\mathcal{M}} &=
  -\bigl(\kappa_d+2\gamma_d+2i\Omega(\tau)\bigr)\mathcal{M}
  +\xi(\tau)\,(2n+1),
  \label{eq:bl_m}
\end{align}
where dots denote $d/d\tau$.

In Equation~\eqref{eq:bl_n} the three terms describe loss, thermal pumping
toward $n_{\rm th}$ and particle creation, the last one driven by the
squeezing $\mathrm{Re}(\mathcal{M})$ in the presence of
non-adiabaticity; dephasing conserves $n$.
In Equation~\eqref{eq:bl_m}, $\mathcal{M}$ rotates at twice the mode
frequency, decays through loss and dephasing, and is sourced by
$\xi(2n+1)$.
This factor follows from \mbox{$[\hat a,\hat a^\dagger]=1$}: the term $2n$
describes stimulated creation, the amplification of photons already
present in the mode (earlier pairs or, for a thermal initial state,
thermal photons), while the $+1$ describes spontaneous creation from
the vacuum, the circuit counterpart of cosmological particle
creation.

As a check, in the absence of dissipation ($\kappa_d=\gamma_d=0$) and starting from the vacuum, $n(\tau_i)=\mathcal{M}(\tau_i)=0$, Equations~\eqref{eq:bl_n}--\eqref{eq:bl_m} reduce to the  unitary Bogoliubov dynamics of Section~\ref{sec:cosmo}. After the pulse the frequency is constant, the instantaneous mode coincides with the out mode, and $n(\tau_f)$ is the number of created particles, $N_k$. For the ideal Weber profile the equations give $N_k=e^{-\pi A}$ and $|\mathcal{M}|=\sqrt{N_k(N_k+1)}$, the exact result of Section \ref{sec:cosmo}; with the finite window of the protocol they reproduce the deviation discussed in Appendix~\ref{app:weber}. This validates the method, which we now apply with $\kappa_d$, $\gamma_d$, $n_{\rm th}\neq0$.

We now include dissipation. As operating point we take the values of Table~\ref{tab:params}. Figure \ref{fig:dissipation} shows $N_k=n(\tau_f)$ and $|\mathcal M(\tau_f)|$ for mode $n=1$, integrating over the modulation window with the mode initially in thermal equilibrium with the bath at its initial frequency, $n(\tau_i)=n_{\rm th}(\omega_k(\tau_i))$, $\mathcal M(\tau_i)=0$. The dashed line is the unitary value of the protocol, $N_k\approx0.60$, used as a reference.
As the loss rate grows, the number of created particles $N_k$ decreases; photon loss is nevertheless mild: for $\kappa_d\lesssim0.01$ both $N_k$ and $|\mathcal M|$ stay within a few percent of their unitary values (Figure~\ref{fig:dissipation}a).
Figure~\ref{fig:dissipation}b shows $N_k$ and $|\mathcal M|$ as functions of the bath temperature. Both remain at their unitary values up to $\approx$50 mK and grow at higher temperature, as the initial thermal photons are amplified by the same Bogoliubov mixing, $N_k\approx N_k^{(0)}+n_{\rm th}(\omega_k(\tau_i))(1+2N_k^{(0)})$; the thermal contribution exceeds the cosmological one only above $T\approx145\,\mathrm{mK}$. The thermal effect is small because the mode is prepared and read out at $\omega_{\max}$, where the bath is essentially in its ground state. The mode reaches low frequencies, where the thermal occupation is larger, only during the brief non-adiabatic crossing, which is too short for the bath to populate it: although $n_{\rm th}\approx0.76$ at the crossing, the thermal photons absorbed during the pulse amount to $\kappa_d\int n_{\rm th}\,d\tau\approx8\times10^{-4}$.
Panels (a) and (b) are computed without dephasing, whose effect is shown in Figure~\ref{fig:dissipation}c. Dephasing leaves $N_k$ essentially unchanged, since it does not change the photon number, but it reduces $|\mathcal M|$, the anomalous moment that measures the squeezing of the mode: by randomising the phase of the mode it destroys the two-photon coherence. The mode is in a quantum, squeezed state, one of its quadratures fluctuating less than in the vacuum, as long as $|\mathcal M|>N_k$. For $\gamma_d\gtrsim0.05$, where the two curves cross, $|\mathcal M|$ falls below $N_k$ and the state is no longer squeezed. It does not become classical, however: dephasing leaves the photon-number distribution unchanged, so the state remains nonclassical, and it keeps a finite $|\mathcal M|$, a trace of the squeezing created by the pulse.

Particle creation and squeezing are two faces of the same process.
The non-adiabatic term $\xi(t)\hat a^\dagger$ of Equation~\eqref{eq:da_dt}
mixes $\hat a$ with $\hat a^\dagger$, which is the action of a squeezing
transformation: without dissipation, the pulse maps the initial vacuum
onto a squeezed vacuum with squeezing parameter $r_k$,
$\tanh r_k=|\beta_k/\alpha_k|$, where $\alpha_k$ and $\beta_k$ are the
Bogoliubov coefficients of Equation~\eqref{eq:bogo}, for which $N_k=\sinh^2r_k$ and
$|\mathcal{M}|=\sinh r_k\cosh r_k=\sqrt{N_k(N_k+1)}$; for the Weber
profile, $\sinh^2r_k=e^{-\pi A}$.
The created particles can therefore be characterised equivalently by
the squeezing of the mode quadratures,
$X_+=(\hat a+\hat a^\dagger)/\sqrt2$ and
$X_-=(\hat a-\hat a^\dagger)/(i\sqrt2)$, with
$\mathrm{Var}(X_\pm)=n+1/2\pm|\mathcal{M}|$.
In the vacuum both variances equal $1/2$; the state is squeezed when
$\mathrm{Var}(X_-)=n+1/2-|\mathcal{M}|<1/2$ (i.e.,\ $|\mathcal{M}|>n$), and the squeezing in dB is
\begin{equation}
  r_{\rm dB}
  = 10\log_{10}\!\left(\frac{1/2}{n+1/2-|\mathcal{M}|}\right).
  \label{eq:sq_dB}
\end{equation}
For the parameters of Table~\ref{tab:params}, Equation~\eqref{eq:sq_dB}
gives $r_{\rm dB}\approx6\,\mathrm{dB}$, and neither loss nor the
$20\,\mathrm{mK}$ thermal background changes it appreciably.
The squeezing disappears only above $T\approx430\,\mathrm{mK}$, where
the amplified thermal photons dominate.

In summary, dissipation does not erase the cosmological signal: for
the parameters of Table~\ref{tab:params}, photon loss, thermal
excitation and dephasing change $N_k$ and the squeezing only slightly,
and the squeezing created by the pulse is comparable to that produced
by Josephson parametric amplifiers~\cite{CastellanosBeltran2008}.
However, two limitations remain.
The photon number cannot distinguish pairs created from the vacuum
from thermal photons amplified by the same mixing; and under strong
dephasing or at high temperature the state is no longer squeezed, although $|\mathcal{M}|$ remains
finite.
This motivates an observable built on $|\mathcal{M}|$, which we
introduce in the following section.

\section{\texorpdfstring{Anomalous Moment}{Anomalous Moment} as a Witness of Bogoliubov Mixing}
\label{sec:correlations}

Section~\ref{sec:dissipation} showed that the photon number $N_k$
cannot distinguish created pairs from thermal photons, and that the
squeezing is lost under strong dephasing or at high temperature, while
$|\mathcal{M}|$ remains finite.
In this section we introduce an observable built on $|\mathcal{M}|$
that vanishes for any thermal state, whether or not the state is
squeezed.
It rests on the pair structure of the Bogoliubov transformation,
Equation~\eqref{eq:bogo}, which mixes modes $k$ and $-k$.
The out-annihilators are
$\hat{b}_k=\alpha_k\hat{a}_k+\beta_k^*\hat{a}_{-k}^\dagger$ and, by
isotropy, $\hat{b}_{-k}=\alpha_k\hat{a}_{-k}+\beta_k^*\hat{a}_{k}^\dagger$:
the in-vacuum is a two-mode squeezed vacuum of the pair $(k,-k)$, with
squeezing parameter $r_k$, $\tanh r_k=|\beta_k/\alpha_k|$, as in
Section~\ref{sec:dissipation}~\cite{Braunstein2005}.
Pairs are created with opposite momenta, as required by momentum
conservation in the homogeneous background.
In the circuit, $k$ and $-k$ are not independent: the open line has
standing-wave modes $u_k\propto\cos(kx)$, with annihilator
$\hat b_c=(\hat b_k+\hat b_{-k})/\sqrt2$ (the sine combination
$\hat b_s=(\hat b_k-\hat b_{-k})/\sqrt2$ violates $\partial_x u|_{x=0}=0$
and is absent), and since
$\langle\hat b_{\pm k}\hat b_{\pm k}\rangle=0$,
$\langle\hat b_c\hat b_c\rangle=\langle\hat b_k\hat b_{-k}\rangle$.
The two-mode squeezing of the pair is therefore the single-mode
squeezing of the circuit mode, described by the $|\mathcal{M}|$ of
Section~\ref{sec:dissipation}.
In the circuit $C_k$ below is therefore a single-mode quantity, and what
it certifies is single-mode squeezing.

We consider the anomalous moment of the out modes, evaluated in
the state $\hat\rho$ of the field at the end of the pulse, as obtained
from the open dynamics of Section~\ref{sec:dissipation}:
\begin{equation}
  C_k \equiv   {\bigl|\langle\hat b_c\hat b_c\rangle-\langle\hat b_c\rangle^2\bigr|},
  \qquad \langle\,\cdot\,\rangle={\rm Tr}[\hat\rho\;\cdot\;],
  \label{eq:Ck_def}
\end{equation}
which for the homogeneous pair reduces to
$|\langle\hat b_k\hat b_{-k}\rangle|$; here and below, the subscript
\mbox{$k=n\pi/D$} labels the standing-wave mode $\cos kx$.
Subtracting the mean fields removes any coherent displacement, so
$C_k$ measures only the noise of the state.
$C_k$ vanishes for every state whose noise is the same in all phase
directions: thermal states at any temperature
($\hat\rho_{\rm th}\propto e^{-\beta\hat H}$ is diagonal in the Fock
basis, so $\hat b_c\hat b_c$ has zero expectation value) and their
displaced versions, including coherent states.
A nonzero $C_k$ thus requires phase-dependent noise.
In the circuit, loss, thermal excitation and dephasing are phase
insensitive, and the only source of phase-dependent noise is the
non-adiabatic term $\xi$ of Equation~\eqref{eq:da_dt}, that is, the
Bogoliubov mixing.
A nonzero $C_k$ therefore shows that Bogoliubov mixing has taken
place, whatever the temperature and the dissipation rates.
The argument rests only on the phase covariance of loss, thermal
excitation and dephasing; closely related moment equations and criteria
were given for cosmological perturbations and for the dynamical Casimir
effect in Refs.~\cite{Campo2005,Busch2014} (see also
Refs.~\cite{Polarski1996,Martin2016} on the quantum origin of cosmological
perturbations).
This is the circuit counterpart of the argument of Lang and
Sch\"utzhold~\cite{Lang2019}, who showed that field correlators are a
more robust witness of cosmological particle creation than the
particle number.
Here, $C_k(\tau_f)=|\mathcal{M}(\tau_f)|$ for the standing-wave mode, so
its behaviour under the three dissipative channels is that of
$|\mathcal{M}|$ in Figure~\ref{fig:dissipation}.

For the unitary evolution from the vacuum the state of the mode
remains pure, and $C_k$ reduces to the value found in
Section~\ref{sec:dissipation},
\begin{equation}
  C_k^{\rm pure} =                  {|\langle{\rm in}|\hat b_c\hat b_c|{\rm in}\rangle|}
                 = |\alpha_k\beta_k^*|
                 = \sqrt{N_k(N_k+1)}=\tfrac12\sinh2r_k,
  \label{eq:Ck_pure}
\end{equation}
so that $C_k$ measures directly how much the vacuum has been squeezed.
The mode starts in a thermal state, for which $C_k=0$, and none of the
dissipative channels can generate a nonzero value: without particle
creation, $C_k$ remains zero throughout the evolution.
Any $C_k\neq0$ at the end of the pulse is therefore produced by the
modulation.

\begin{figure}[h!]
  \begin{minipage}[b]{0.48\textwidth}\centering
  \includegraphics[width=\linewidth]{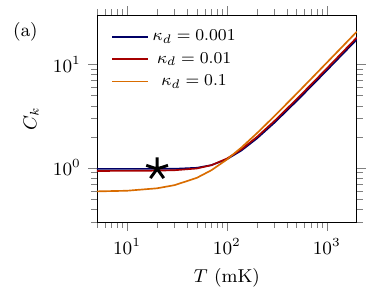}\end{minipage}\hfill
  \begin{minipage}[b]{0.48\textwidth}\centering
  \includegraphics[width=\linewidth]{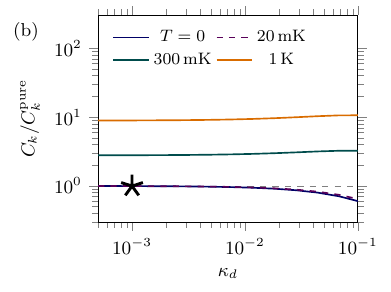}\end{minipage}\\[4pt]
  \begin{minipage}[b]{0.48\textwidth}\centering
  \includegraphics[width=\linewidth]{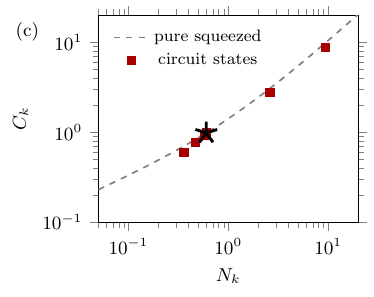}\end{minipage}\hfill
  \begin{minipage}[b]{0.48\textwidth}\centering
  \includegraphics[width=\linewidth]{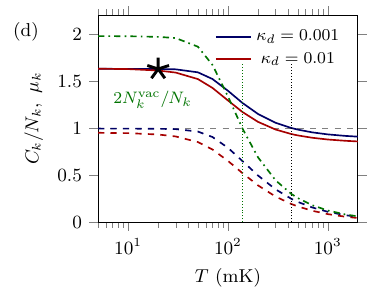}\end{minipage}
  \caption{%
    The anomalous
 moment $C_k=|\mathcal{M}(\tau_f)|$ for mode $n=1$,
    for the parameters of Table~\ref{tab:params}, from the open dynamics
     with a thermal initial
    state. The star marks the operating point.
    (\textbf{a}) $C_k$ vs. $T$ for three values of $\kappa_d$: $C_k>0$ at all
    temperatures, whereas any thermal state has $C_k=0$.
    \mbox{(\textbf{b}) Ratio} $C_k/C_k^{\rm pure}$ to the value for the unitary
    evolution from the vacuum, Equation~\eqref{eq:Ck_pure}, vs. $\kappa_d$ for
    four temperatures; at high $T$ it starts at $2n_{\rm th}+1$, the
    amplification factor of the initial thermal photons.
    \mbox{(\textbf{c}) States} in the $(N_k,C_k)$ plane; the dashed curve
    $C_k=\sqrt{N_k(N_k+1)}$ contains the pure states. Squares: loss only
    ($n_{\rm th}=0$) and temperature at fixed $\kappa_d=0.001$.
    (\textbf{d}) $C_k/N_k$ (solid) and purity $\mu_k$ (dashed) vs. $T$ for
    $\kappa_d=0.001$ and $0.01$; above the dashed line $C_k/N_k=1$ the
    mode is squeezed [Equation~\eqref{eq:entangled}], and the dotted line
    marks the crossing for $\kappa_d=0.001$; green: vacuum-dominance ratio $2N_k^{\rm vac}/N_k$ for $\kappa_d=0.001$, which crosses unity at $\approx140$ mK (green dotted line).
  }
  \label{fig:correlations}
\end{figure}

A nonzero $C_k$ at the end of the pulse shows that the evolution has
been non-adiabatic, that is, that the squeezing term $\xi$ has left a
net Bogoliubov mixing ($\beta_k\neq0$).
Such mixing always creates pairs from the vacuum, which contribute
$\sinh^2r_k$ to $N_k$.
If the mode starts in a thermal state, the same mixing also amplifies
the thermal photons, which add $n_{\rm th}\cosh2r_k$ to $N_k$ and
contribute to $C_k$ as well.
$C_k$ alone does not tell which of the two contributions dominates;
the purity of the state and the criterion below do.
For the Gaussian state of the standing-wave mode $\hat b_c$ the purity
is
\begin{equation}
  \mu_k = \frac{1}{2\sqrt{\bigl(N_k+\tfrac12\bigr)^2-C_k^2}},
  \label{eq:purity}
\end{equation}
where $N_k=n(\tau_f)$ is the photon number at the end of the pulse.
For the pure state created from the vacuum, $C_k^2=N_k(N_k+1)$ and
$\mu_k=1$.
Bogoliubov mixing preserves the purity, so a thermal initial state
gives $\mu_k=1/(2n_{\rm th}+1)$, and dissipation lowers $\mu_k$
further.
The quantum character of the pairs is certified by
\begin{equation}
  C_k > N_k,
  \label{eq:entangled}
\end{equation}
which  in the standing-wave mode of
the circuit is equivalent to squeezing,
\mbox{$N_k+1/2-C_k<1/2$} [Equation~\eqref{eq:sq_dB}].
Three regimes follow: $C_k=0$ (no Bogoliubov mixing); $0<C_k\leq N_k$
(Bogoliubov mixing, but no squeezing; for the Gaussian states produced by loss and thermal noise, the state is classical); and $C_k>N_k$ ( the state is squeezed and quantum).
Squeezing does not mean that the pairs created from the vacuum
dominate. Since the mixing preserves the purity,
$N_k+\tfrac12=(n_{\rm th}+\tfrac12)\cosh2r_k$ and $\mu_k=1/(2n_{\rm th}+1)$,
so the vacuum contribution,
$N_k^{\rm vac}=\sinh^2r_k=\mu_k(N_k+\tfrac12)-\tfrac12$, follows from $N_k$
and $C_k$. It dominates for $N_k^{\rm vac}>N_k/2$, a stricter condition:
for $n_{\rm th}=0.5$ and $r_k=0.4$, $C_k=0.89>N_k=0.84$ although the
thermal part is four times the vacuum part.
This estimate assumes no loss between the pulse and the detector. A detection efficiency $\eta$ multiplies $N_k$ and $C_k$ by $\eta$, which leaves $C_k/N_k$ unchanged but lowers the purity: for the pure squeezed vacuum of the operating point, $N_k^{\rm vac}>N_k/2$ fails for $\eta\lesssim0.58$.
Figure~\ref{fig:correlations} illustrates the three roles of $C_k$ for mode $n=1$.

\textit{$C_k$ as a witness.
} Panel (a) shows $C_k$ as a function of the bath temperature for three loss rates. $C_k>0$ at all temperatures, whereas any thermal state has $C_k=0$. Below $\approx50$ mK the curves are flat, and at the operating point (star) $C_k\approx0.98$ is essentially its unitary value. Above $\approx100$ mK $C_k$ grows with $T$. This does not mean that more pairs are created from the vacuum: the thermal photons present at the start are amplified by the same mixing. Panel (b) isolates the effect of loss through the ratio $C_k/C_k^{\rm pure}$, with $C_k^{\rm pure}$ the value for the unitary evolution from the vacuum (Equation~\eqref{eq:Ck_pure}). Loss reduces $C_k$ smoothly and in the same way at all temperatures, while at high temperature the ratio starts at $2n_{\rm th}+1$, the amplification factor of the initial thermal photons.

\textit{$C_k$ as a measure of the squeezing.} Panel (c) places each state in the plane $(N_k,C_k)$. The pure states lie on the curve $C_k=\sqrt{N_k(N_k+1)}$ (dashed), where $C_k$ is fixed by the squeezing created by the pulse; the operating point (star) lies on it. Loss and temperature move the states below this curve, towards mixed states, but never onto $C_k=0$.

\textit{Quantum character.} Panel (d) shows $C_k/N_k$ and the purity $\mu_k$ as functions of temperature. Up to $\approx50$ mK, $\mu_k\approx1$ and $C_k/N_k>1$: the state is the pure squeezed vacuum, and the pairs come from the vacuum. At higher temperature the amplified thermal photons increase $N_k$ faster than $C_k$ and lower the purity towards $1/(2n_{\rm th}+1)$. $C_k/N_k$ crosses unity at $\approx430$ mK (black dotted line). Below this temperature the mode is squeezed [Equation~\eqref{eq:entangled}]; above it the state is classical, although $C_k$ remains nonzero. The vacuum contribution $N_k^{\rm vac}$ stays at $0.58$--$0.59$ at all these temperatures and dominates only below $\approx140$ mK (green curve in panel (d)). Loss barely changes $C_k/N_k$ but lowers $\mu_k$. This is why, for a mixed state, the purity alone cannot establish the quantum character, and the criterion $C_k>N_k$ \mbox{is needed}.

These results were obtained for mode $n=1$ of the line of
Table~\ref{tab:params}. The same ideal dynamics is obtained for a
family of circuits with a different frequency scale, which, however,
differ in their sensitivity to temperature and to timing errors
(Appendix~\ref{app:validity}).

\section{Conclusions}
\label{sec:conclusions}

We have proposed an analogue of cosmological particle creation, together with its error budget, on the
circuit-QED platform of Ref.~\cite{Tian2017}: a superconducting transmission line with
distributed dc-SQUIDs, whose modes reproduce a scalar field in a
radiation-dominated FRW universe in conformal time.
With a suitable flux protocol, each mode of the line obeys the same
Weber equation as the cosmological model, so that the in-vacuum evolves
into a squeezed state with the cosmological Bogoliubov spectrum.
The fidelity of the analogue requires a pulse long enough to reproduce
the Weber spectrum, a uniform modulation that keeps the Bogoliubov
transformation diagonal, and nearly symmetric SQUIDs; since these
requirements favour different pulse lengths, the design results from a
compromise between them.
Once this compromise is fixed, the same ideal dynamics is obtained for
circuits with different frequency scales, which differ in their
sensitivity to temperature and to timing errors.
Dissipation does not erase the cosmological signal: photon loss and
thermal excitation barely change it, and the squeezing created by the
pulse is comparable to that of Josephson parametric amplifiers.

We used the anomalous moment $C_k$ as a witness of the
Bogoliubov mixing.
It vanishes for any thermal state and for an adiabatic evolution, so a
nonzero $C_k$ shows that the pulse has created pairs, and it survives
dissipation even when strong dephasing or high temperature remove the squeezing.
For the unitary evolution from the vacuum its value measures the
squeezing created by the pulse.
Together with the photon number, it
tells whether the mode is squeezed, and with the purity whether the
pairs created from the vacuum dominate.

In any physical realisation of the analogue, the transmission line
would be cooled to millikelvin temperatures with the flux set so that every mode sits at its highest frequency, where it would be thermally empty and start in the vacuum. The flux pulse, delivered simultaneously to all SQUIDs, would drive the modes through the non-adiabatic crossing and return them to the same frequency, where they could be measured by homodyne detection before they thermalise.
The mode $n=0$ ($A=0$) is the most populated, $N_0\approx0.95$, but at
the readout frequency it lies $13\,\mathrm{MHz}$ below mode $n=1$, against
$\kappa/2\pi=1\,\mathrm{MHz}$; a temporal filter orthogonal to it changes
the squeezing of mode $n=1$ from $6.1$ to $6.0\,\mathrm{dB}$. A control run with the modulation switched off, which should give $C_k=0$, would rule out squeezing generated in the readout chain.

Switching the Josephson term on and off smoothly removes the error of the abrupt edges, so that the first four modes test the Gaussian dependence on $k$, although for $n=3,4$ this requires tighter tolerances on the asymmetry, the flux and the timing than for $n=1$ (Appendix~\ref{app:validity}).

Natural extensions include probing other cosmological backgrounds
through different time profiles of the Josephson inductance; measuring
the two-point correlator of the field along the line to access the
light-cone signatures of pair creation~\cite{Lang2019}; and exploiting
the analogy with the Schwinger effect to probe strong-field QED
physics.

\begin{acknowledgments}
We are grateful to F.D. Mazzitelli for relevant and constructive comments in the early stages of this manuscript. This research was funded by Consejo Nacional de Investigaciones Cient\'ificas y T\'ecnicas (CONICET), and Universidad de Buenos Aires (UBA).
\end{acknowledgments}

\appendix

\section{Fidelity of the Circuit Analogue}
\label{app:validity}
\label{app:params}

In this section we review the considerations that enter the design of
the \mbox{analogue circuit.}

The fidelity of the circuit analogue results from a compromise between
three main requirements: (i) the validity of the Weber prediction
despite the finite duration of the pulse, (ii) the robustness of the signal
against SQUID asymmetry, and (iii) a nearly diagonal
Bogoliubov transformation; in addition, (iv) the circuit must remain
linear.
The design is fixed by three choices: the frequency $\omega_{k0}$ of
the mode, its Weber parameter $A$, and
$\tau_*^2=\omega_{\max}^2/m_{\rm eff}$.
They determine
\begin{equation}
  m_{\rm eff}=\frac{\omega_{k0}^2}{A},\qquad
  t_*=\frac{\tau_*}{\sqrt{m_{\rm eff}}},\qquad
  \omega_{\max}=\tau_*\sqrt{m_{\rm eff}},\qquad
  I_c=\frac{\Phi_0\,C\,\omega_{\max}^2}{2\pi}\propto\omega_{k0}^2,
\end{equation}
for a given shunt capacitance $C$; for a given length $D$, the mode
number fixes the propagation speed $v$.

\begin{figure}[h!]
  \begin{minipage}[b]{0.48\textwidth}\centering
  \includegraphics[width=\linewidth]{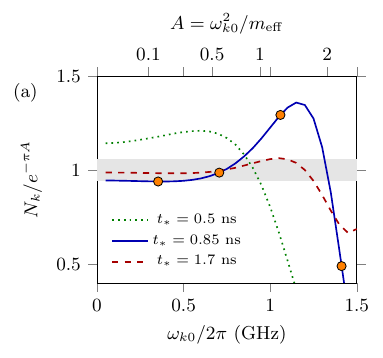}\end{minipage}\hfill
  \begin{minipage}[b]{0.48\textwidth}\centering
  \includegraphics[width=\linewidth]{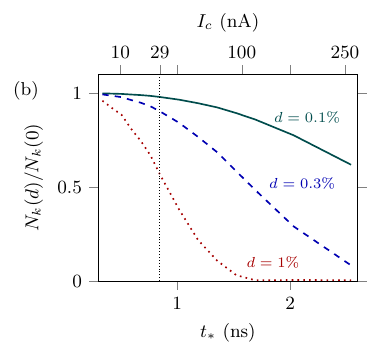}\end{minipage}
  \caption{%
    Finite duration of the pulse and SQUID asymmetry, from the exact
    protocol with vacuum initial state and no dissipation, in a
    uniform line.
    (\textbf{a}) Ratio $N_k/e^{-\pi A}$ vs. mode frequency $\omega_{k0}$ for a
    fixed pulse ($\sqrt{m_{\rm eff}}/2\pi=0.94\,\mathrm{GHz}$, so
    $A=\omega_{k0}^2/m_{\rm eff}$ grows with $\omega_{k0}$, top axis):
    $\tau_*=3$ ($t_*=0.5\,\mathrm{ns}$, $I_c=10\,\mathrm{nA}$, dotted),
    $\tau_*=5$ ($t_*=0.85\,\mathrm{ns}$, $I_c=29\,\mathrm{nA}$, this
    work, solid) and $\tau_*=10$ ($t_*=1.7\,\mathrm{ns}$,
    $I_c=115\,\mathrm{nA}$, dashed); the gray band marks $\pm6\%$;
    circles: modes $n=1$--$4$ of the line.
    (\textbf{b}) $N_k$ of mode $n=1$ ($A=0.14$) relative to symmetric SQUIDs,
    $N_k(d)/N_k(0)$, vs. pulse length $t_*$ at fixed $m_{\rm eff}$
    (top axis: corresponding $I_c$), for asymmetries $d=0.1\%$,
    $0.3\%$ and $1\%$; the dotted line marks this work.
  }
  \label{fig:edge}
\end{figure}

\textit{(i) Finite duration of the pulse.}
\label{app:weber}
The analogue is defined by a single pulse applied to all SQUIDs of the
line.
Ideally, the frequency of every mode would follow
$\omega_k^2=\omega_{k0}^2+m_{\rm eff}^2t^2$ at all times, and every mode
would give the Weber result $N_k=e^{-\pi A}$.
In the circuit, however, the modulation lasts only $|t|\leq t_*$ and
stops abruptly, so $N_k$ can differ from $e^{-\pi A}$: for mode $n=1$
it is $6\%$ lower for $\tau_*=5$, and more than $8\%$ off for
$\tau_*\lesssim4$, where the abrupt end of the pulse creates additional
particles.
The deviation depends on the frequency of the mode, $\omega_{k0}$, the
minimum it reaches at $t=0$, which is fixed by the parameters $D$,
$L_0$ and $C$ of the line.
For a given pulse (fixed $t_*$ and $m_{\rm eff}$, hence fixed
$\tau_*$), a higher $\omega_{k0}$ means a larger
$A=\omega_{k0}^2/m_{\rm eff}$ (Figure~\ref{fig:edge}).
The edges of the pulse create a small, nearly $A$-independent
amplitude of pairs, which interferes with the one created at the
non-adiabatic crossing.
While $e^{-\pi A}$ is large this gives the constant $-6\%$; when
$e^{-\pi A}$ becomes small the edge contribution is comparable and
$N_k$ departs strongly from the Weber result.
For the pulse of Table~\ref{tab:params}, $N_k$ follows $e^{-\pi A}$
within $6\%$ up to $\omega_{k0}/2\pi\approx0.8\,\mathrm{GHz}$
($A\approx0.7$), which includes modes $n=1,2$, whereas modes $n=3$
and $4$ ($1.06$ and $1.41\,\mathrm{GHz}$) deviate by $+29\%$ \mbox{and
$-51\%$.}

A longer pulse extends the faithful range, to
$\approx1.2\,\mathrm{GHz}$ for $\tau_*=10$.
A smooth switching does better: with
$1/[L_J(t)C]=m_{\rm eff}^2t_*^2\tanh(t^2/t_*^2)$, which coincides with
the Weber profile near the crossing, modes $n=1$--$4$ follow
$e^{-\pi A}$ within $0.3\%$, at the cost of a pulse about twice as long;
the sensitivity to skew and asymmetry is unchanged.

\textit{(ii) Robustness against SQUID asymmetry.}
A longer pulse cannot be chosen freely, however.
At fixed $m_{\rm eff}$ it requires a larger $I_c$, since
$\omega_{\max}=\tau_*\sqrt{m_{\rm eff}}$.
For asymmetric SQUIDs the Josephson term does not vanish at
$\Phi_0/2$: a residual $d\,E_J^{\max}$ remains; near the crossing
$\Omega^2\approx A+(\tau^4+d^2\tau_*^4)^{1/2}$, whose minimum is
$A+d\,\tau_*^2$ and which is no longer of Weber form once
$d\,\tau_*^2\gtrsim A$. This suppresses pair creation (all results use
the exact expression).
The same asymmetry therefore weighs more the longer the pulse
(Figure~\ref{fig:edge}b).


\textit{(iii) Mode mixing.}\label{app:mixing}
The results of (i) and (ii)  assume a perfectly uniform line, in which every
cell sees the same Josephson term and each mode evolves independently.
In a real circuit the cells differ slightly: the flux pulse reaches
them at slightly different times, and their critical currents and
asymmetries are not identical.
These differences couple different modes, so that pairs are created
not only between $k$ and $-k$ but also between modes $n\neq m$
($\beta_{nm}\neq0$), the Bogoliubov transformation is no longer
diagonal, and the correspondence with the homogeneous cosmological
model is lost.
Since the differences are fractions of the Josephson term, which
reaches $\omega_{\max}^2=\tau_*^2m_{\rm eff}$, while the modes are
separated by $\sim$$\omega_{k0}^2$, the mixing grows with $\tau_*^2$, that
is, with the length of the pulse.
We measure this by the fraction of the pairs of mode $n=1$ created
with other modes,
$f_1=\sum_{m\neq1}|\beta_{1m}|^2/\sum_m|\beta_{1m}|^2$, which vanishes
for a diagonal transformation.
Table~\ref{tab:improve} gives $f_1$ and $N_k$ for mode $n=1$ for
pulses of different length at fixed $m_{\rm eff}$, with
representative inhomogeneities of the modulation (an arrival-time difference
$\Delta t=3\,\mathrm{ps}$ between the ends of the line, a random timing
error of $3\,\mathrm{ps}$ per cell, a $1\%$ spread of $I_c$ and
$d=(1\pm0.5)\%$), from the coupled equations of the modes $\cos(kx)$
with a cell-dependent Josephson term ($10^3$ cells).
An arrival-time difference of a few ps requires a bias network with
equal path lengths to all SQUIDs: a single feed line gives
$\Delta t\approx D/v_b\approx0.5\,\mathrm{ns}$ ($v_b\approx1.2\times10^8\,\mathrm{m/s}$),
for which $f_1\approx0.95$. Table~\ref{tab:skew} gives $N_1$ and $f_1$ as
functions of the skew. The equal-path values assume independent residual delays in each cell. In a real distribution network, such as an H-tree, the cells of one branch share its delay error, and the mixing is driven by the long-wavelength part of the delay profile: a branch-correlated profile with the same rms gives $f_1\approx0.05$ at $10\,\mathrm{ps}$ and $0.36$ at $30\,\mathrm{ps}$, close to the single feed. The tolerance then drops to a few ps: $f_1<1\%$ requires the lowest cosine components of the delay profile to stay below $\approx$$1.5$--$3\,\mathrm{ps}$. The peak flux must also reach $\Phi_0/2$ within
$\sim10^{-4}\Phi_0$: an error of $10^{-3}\Phi_0$ changes $N_1$ by $-22\%$
(below $\Phi_0/2$) or $+12\%$ (above). A finite bandwidth of the flux
line acts mainly through this peak deficit: for a $5\,\mathrm{GHz}$
low-pass filter $N_1$ drops by $5$--$9\%$, and by $\lesssim$$2\%$ once the
peak is recalibrated.
A static spread of the applied flux between cells (loop areas, mutual inductances) matters near $\Phi_0/2$, where $|\cos\pi\Phi/\Phi_0|$ is rectified and the errors do not average out: an independent spread of $10^{-3}\Phi_0$ lowers $N_1$ by $6\%$, and $3\times10^{-3}\Phi_0$ by $28\%$; a $1\%$ spread of the loop area, $5\times10^{-3}\Phi_0$ at the peak, lowers it by $51\%$.
These tolerances are for mode $n=1$. Modes $n=3$ and $4$, whose $N_k$ is one and three orders of magnitude smaller, are more sensitive. With smooth switching, $d=0.3\%$ lowers $N_3$ and $N_4$ by $24\%$ and $60\%$, and the mixing transfers pairs from the more populated low modes: with $d=0.1\%$ and branch-correlated delays of $10\,\mathrm{ps}$, the total photon number of modes $3$ and $4$ rises by $\approx$$25\%$ and $\approx$$140\%$. Keeping both within $\approx$$15\%$ requires $d\lesssim0.05\%$, correlated \mbox{delays $\lesssim3\,\mathrm{ps}$} and a flux \mbox{spread $\lesssim3\times10^{-4}\Phi_0$.}
\begin{table}[t]
  \caption{Mode $n=1$ vs. arrival-time skew of the flux pulse ($\tau_*=5$,
  $N_c=10^3$, symmetric SQUIDs): single feed at one end (linear
  delay, $\Delta t$ end to end) and equal-path network with random
  residual delays of standard deviation $\Delta t$, independent in each cell. $N_1=|\beta_{11}|^2$ is the diagonal part of the photon number of mode $1$, whose total changes much less with the skew.}
  \label{tab:skew}
  \begin{tabular}{lcccc}
    \toprule
    \boldmath{$\Delta t$} & \boldmath{$3\,\mathrm{ps}$} & \boldmath{$10\,\mathrm{ps}$} & \boldmath{$15\,\mathrm{ps}$} & \boldmath{$30\,\mathrm{ps}$} \\
    \midrule
    Single feed: $f_1$ & $0.004$ & $0.045$ & $0.10$ & $0.36$ \\
    Single feed: $N_1$ & $0.60$ & $0.58$ & $0.54$ & $0.39$ \\
    Equal paths: $f_1$ & $8\times10^{-5}$ & $9\times10^{-4}$ & $2\times10^{-3}$ & $8\times10^{-3}$ \\
    Equal paths: $N_1$ & $0.60$ & $0.60$ & $0.60$ & $0.57$ \\
    \bottomrule
  \end{tabular}
\end{table}

\textit{(iv) Linearity of the circuit.}\label{app:phase}
The quadratic expansion of the Josephson energy requires small
node-flux fluctuations, $\phi_n^2\equiv(2\pi\Phi_n/\Phi_0)^2\ll1$.
The zero-point fluctuations of the node flux, summed over all modes
of a line of $N_c$ cells with lattice dispersion, including the mode
$n=0$ and the growth of the low modes during the pulse, give
\mbox{$\langle\phi_n^2\rangle\approx0.02$--$0.03$} for $N_c=10^3$ (at most $0.034$, of
which the mode $n=0$ contributes $0.004$), so that the
quartic (Kerr) correction relative to the quadratic term,
$\sim\langle\phi_n^2\rangle/12$, is $\lesssim0.3\%$; the pairs created
($N_k\lesssim1$ per mode, spread over $N_c$ nodes) do not change this.
Phase slips are excluded because each node is bound to its neighbours
by the inductive coupling through $L_0=a^2/(v^2C)$, which dominates
over the Josephson term ($L_0/L_{J,\min}\approx2\times10^{-3}$ for
$N_c=10^3$), so the potential of each node has a single well.
The reduced $E_J^{\max}/E_C\approx75$ of the weaker SQUIDs is
therefore not a limitation.
For $N_c=10^2$ the Kerr correction rises to $\approx2\%$ and
$L_0/L_{J,\min}\approx0.2$, which sets a practical lower bound of
order $10^3$ cells for $D=6\,\mathrm{cm}$.

\begin{table}[t]
  \caption{Pulse length
 and fidelity of the analogue for mode $n=1$
  ($A=0.14$, fixed $m_{\rm eff}$), with representative
  inhomogeneities of the modulation, from the coupled-mode equations ($N_c=10^3$, three
  realisations). The deviation from the Weber result is for symmetric
  SQUIDs.}
  \label{tab:improve}

  \begin{tabular}{lcccc}
    \toprule
     & \boldmath{$t_*=0.5\,\mathrm{ns}$} & \boldmath{$t_*=0.85\,\mathrm{ns}$} \textbf{(This Work)} & \boldmath{$t_*=1.2\,\mathrm{ns}$} & \boldmath{$t_*=1.7\,\mathrm{ns}$} \\
    \midrule
    $\tau_*$ & $3$ & $5$ & $7$ & $10$ \\
    $I_c$ & $10\,\mathrm{nA}$ & $29\,\mathrm{nA}$ & $56\,\mathrm{nA}$ & $115\,\mathrm{nA}$ \\
    \midrule
    Deviation from $e^{-\pi A}$ & $+18\%$ & $-6\%$ & $-2.5\%$ & $-2\%$ \\
    Mode mixing $f_1$ & $1.6\times10^{-3}$ & $4.7\times10^{-3}$ & $1.0\times10^{-2}$ & $4.6\times10^{-2}$ \\
    $N_k$, symmetric SQUIDs & $0.76$ & $0.60$ & $0.63$ & $0.63$ \\
    $N_k$, $d=(1\pm0.5)\%$ & $0.66$ & $0.33$ & $0.13$ & $0.003$ \\
    \bottomrule
  \end{tabular}

\end{table}

\begin{figure}[h!]
  \includegraphics[width=0.85\textwidth]{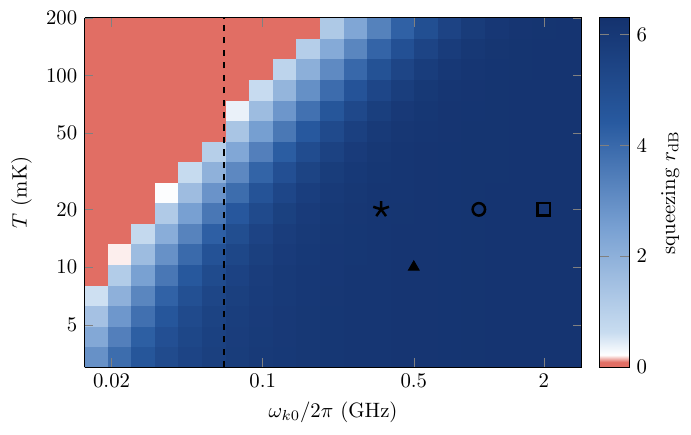}
  \caption{%
    Squeezing of circuits with the same ideal dynamics in the
    frequency--temperature plane ($A=0.14$, $\tau_*=5$, $\kappa/2\pi=1\,\mathrm{MHz}$): each
    frequency corresponds to a different line, whose mode $n=1$ has
    frequency $\omega_{k0}$, driven by a rescaled pulse,
    $t_*=\tau_*\sqrt{A}/\omega_{k0}$.
    Red: no squeezing. The map ends at $3\,\mathrm{GHz}$, where
    $t_*=0.1\,\mathrm{ns}$. Dashed line: $I_c=1\,\mathrm{nA}$.
    Star: operating point (Table~\ref{tab:params}). Other points:
    $500\,\mathrm{MHz}$ at $10\,\mathrm{mK}$ (triangle), $1\,\mathrm{GHz}$
    (circle) and $2\,\mathrm{GHz}$ (square) at $20\,\mathrm{mK}$.
  }
  \label{fig:viability}
\end{figure}

\textit{(v) Frequency scale.}\label{app:equiv}
Of the three choices, only $A$ and $\tau_*$ enter the dynamics: in
dimensionless time the mode equation depends on them alone.
For fixed $A$ and $\tau_*$, a change of $\omega_{k0}$ therefore defines
a circuit with the same ideal dynamics, a different line whose mode $n=1$ has frequency
$\omega_{k0}$, driven by a rescaled pulse, $t_*=\tau_*\sqrt{A}/\omega_{k0}$,
with $I_c\propto\omega_{k0}^2$; it gives the same $N_k$, $C_k$ and
squeezing.
These circuits differ only in what does not rescale.
The thermal population of the mode is set by $\hbar\omega_{\max}/k_BT$,
since the bath temperature is fixed.
Figure~\ref{fig:viability} shows the squeezing of these
circuits in the frequency--temperature plane: it takes the value of
the operating point wherever $\hbar\omega_{\max}\gg k_BT$, and is lost
along a line $T\propto\omega_{k0}$.
The timing errors of the modulation, in turn, enter through
$\sqrt{m_{\rm eff}}\,\Delta t$, so a lower frequency tolerates larger
errors: for the same imperfections, $f_1$ is about seven times larger
at $1\,\mathrm{GHz}$ than at $0.35\,\mathrm{GHz}$.
This favours the relatively low frequency of mode $n=1$ adopted here.

Requirements (i)--(iii) imply different pulse lengths.
Short pulses keep the mixing small and are robust against asymmetry,
but depart from the Weber result; long pulses are faithful to it, but
the modes mix more and the asymmetry suppresses pair creation.
The pulse adopted here, $\tau_*=5$, balances the three.
Table~\ref{tab:budget} collects the resulting error budget: the
tolerance on each imperfection and its effect on the signal of mode
$n=1$.
\begin{table}[t]
  \caption{Error budget for mode $n=1$ ($A=0.14$, $\tau_*=5$, $N_c=10^3$;
  $N_1=0.60$ for the ideal protocol).}
  \label{tab:budget}

  \begin{tabular}{llll}
    \toprule
    \textbf{Imperfection} & \textbf{Tolerance} & \textbf{Effect on Mode} \boldmath{$n=1$} & \textbf{Section} \\
    \midrule
    Skew, equal paths, independent & $\Delta t\lesssim30\,\mathrm{ps}$ & $f_1<10^{-2}$, $N_1$: $\leq6\%$ & A
 (iii) \\
    Skew, equal paths, correlated & {$\Delta t\lesssim3\,\mathrm{ps}$} & {$f_1<10^{-2}$, $N_1$: $<1\%$} & App.~A\,(iii) \\
    Skew, single feed & excluded & $f_1\approx0.95$ & App.~A\,(iii) \\
    Peak flux error & $\lesssim$$10^{-4}\Phi_0$ & $N_1$: $\pm2\%$ ($-22\%/{+12\%}$ at $10^{-3}\Phi_0$) & App.~A\,(iii) \\
    Flux-line bandwidth & $\gtrsim$$5\,\mathrm{GHz}$ & $N_1$: $\leq2\%$, peak recalibrated & App.~A\,(iii) \\
    SQUID asymmetry & $d\lesssim0.1$--$0.3\%$ & $N_1$: $-2$ to $-10\%$ ($-43\%$ at $d=1\%$) & App.~A\,(ii) \\
    Flux spread & {$\lesssim$$10^{-3}\Phi_0$} & {$N_1$: $-6\%$ ($-28\%$ at $3\times10^{-3}\Phi_0$)} & App.~A\,(iii) \\
    Spread of $I_c$ & $\lesssim$$3\%$ & $f_1\leq3\times10^{-3}$, $N_1$: $<1\%$ & App.~A\,(iii) \\
    Abrupt pulse edges & smooth switching & $N_1$: $-6\%\to<0.3\%$ ($n\leq4$) & App.~A\,(i) \\
    Mode $n=0$ at readout & $13\,\mathrm{MHz}\gg\kappa/2\pi$ & squeezing $6.1\to6.0\,\mathrm{dB}$ & Section~\ref{sec:conclusions} \\
    Kerr nonlinearity & $N_c\gtrsim10^3$ & $\lesssim$$0.3\%$ & App.~A\,(iv) \\
    Loss, temperature & $1\,\mathrm{MHz}$, $20\,\mathrm{mK}$ & $N_1$: $-0.3\%$ & Section~\ref{sec:dissipation} \\
    \bottomrule
  \end{tabular}

\end{table}
 All in all, Table~\ref{tab:params} collects the circuit parameters of this work, all of them typical of current circuit-QED devices~\cite{Krantz2019}.

\begin{table}[t]
  \caption{Circuit parameters
 of this work: chosen values and derived
  quantities.}
  \label{tab:params}
  \begin{tabular}{lc}
    \toprule
    \multicolumn{2}{l}{\textit{Chosen values}} \\
    Mode $n=1$: operating frequency (preparation and readout), & \\
    \quad $\sqrt{\omega_{k0}^2+\omega_{\max}^2}/2\pi\approx\omega_{\max}/2\pi$ & $4.7\,\mathrm{GHz}$ \\
    Mode $n=1$: minimum frequency $\omega_{k0}/2\pi$ (at $t=0$) & $0.35\,\mathrm{GHz}$ \\
    Mode $n=1$: Weber parameter $A$ & $0.14$ \\
    $\tau_*^2=\omega_{\max}^2/m_{\rm eff}$ & $25$ \\
    Line length $D$ & $6\,\mathrm{cm}$ \\
    Number of cells $N_c$ & $10^3$ \\
    Shunt capacitance per cell $C$ & $0.1\,\mathrm{pF}$ \\
    Loss rate $\kappa/2\pi$ (at the readout frequency) & $1\,\mathrm{MHz}$ \\
    Temperature $T$ & $20\,\mathrm{mK}$ \\
    \midrule
    \multicolumn{2}{l}{\textit{Derived quantities}} \\
    Effective mass $\sqrt{m_{\rm eff}}/2\pi$ & $0.94\,\mathrm{GHz}$ \\
    Modulation window $t_*$ & $0.85\,\mathrm{ns}$ \\
    SQUID critical current $I_c$ & $29\,\mathrm{nA}$ \\
    Minimum Josephson inductance $L_{J,\min}=\Phi_0/(2\pi I_c)$ & $11\,\mathrm{nH}$ \\
    Propagation speed $v=\omega_{k0}D/\pi$ & $4.2\times10^{7}\,\mathrm{m/s}$ \\
    Cell size $a=D/N_c$ & $60\,\mu\mathrm{m}$ \\
    Line inductance per cell $L_0=a^2/(v^2C)$ & $20\,\mathrm{pH}$ \\
    Modes $n=2,3$: $\omega_{k0}/2\pi$ & $0.71,\ 1.06\,\mathrm{GHz}$ \\
    Modes $n=2,3$: $A$ & $0.57,\ 1.27$ \\
    Dimensionless loss rate $\kappa_d=\kappa/\sqrt{m_{\rm eff}}$ & $10^{-3}$ \\
    \bottomrule
  \end{tabular}
\end{table}

\section{Validity of the Lindblad Description}
\label{app:lindblad}

The master equation~\eqref{eq:lindblad}, with jump operators defined
in the instantaneous basis and rates evaluated at the instantaneous
frequency, is the adiabatic Markovian master equation~\cite{Albash2012}.
Its derivation requires weak coupling ($\kappa/\omega_k\lesssim3\times10^{-3}$
here), a bath correlation time short compared with the relaxation time
($\tau_B\sim\hbar/k_BT\approx0.4\,\mathrm{ns}$ at $20\,\mathrm{mK}$,
against $1/\kappa\approx160\,\mathrm{ns}$ for
$\kappa/2\pi=1\,\mathrm{MHz}$), and an adiabatic evolution of the
system on the scale $\tau_B$.
The last condition does not hold during the pulse: the mode goes
through the non-adiabatic crossing, where $|\dot\Omega/\Omega^2|$
is of order unity ($|\tau|\lesssim1$), and the whole modulation lasts
$2t_*\approx1.7\,\mathrm{ns}\approx4\tau_B$ (less than $\tau_B$ for the
shorter pulses of Figure~\ref{fig:viability}).
Equation~\eqref{eq:lindblad} is therefore not derived for the
evolution considered here; we use it as an effective description.
The approximation is nevertheless controlled.
The non-adiabatic interval is short compared with $1/\kappa$; the
Hamiltonian part of Equation~\eqref{eq:lindblad} contains the full $H_k(t)$,
so the non-adiabatic evolution that creates the pairs is treated
exactly, and the approximation affects only the dissipator, whose
total effect during the pulse is of order
$\kappa_d\,2\tau_*\,(n_{\rm th}+1)\sim10^{-2}$ at the operating point.
The description is therefore controlled for $\kappa_d\,2\tau_*\ll1$,
which includes the operating point and the loss rates of current
circuit-QED devices; the largest value shown in
Figures~\ref{fig:dissipation} and~\ref{fig:correlations},
$\kappa_d=0.1$, is an extreme case outside this range.



\begin{thebibliography}{99}
\bibitem{Parker1969}
Parker, L. Quantized fields and particle creation in expanding universes.
\textit{Phys.\ Rev.}\ \textbf{1969}, \emph{183}, 1057.

\bibitem{Birrell1982}
Birrell, N.D.; Davies, P.C.W.
\textit{Quantum Fields in Curved Space};
Cambridge University Press: Cambridge, UK, 1982.

\bibitem{Ford2021}
Ford, L.H. Cosmological particle production: A review.
\textit{Rep.\ Prog.\ Phys.}\ \textbf{2021}, \textit{84}, 116901.

\bibitem{Mukhanov1992}
Mukhanov, V.F.; Feldman, H.A.; Brandenberger, R.H.
Theory of cosmological perturbations.
\textit{Phys.\ Rep.}\ \textbf{1992}, \textit{215}, 203.

\bibitem{Unruh1981}
Unruh, W.G. Experimental black-hole evaporation?
\textit{Phys.\ Rev.\ Lett.}\ \textbf{1981}, \textit{46}, 1351.

\bibitem{Fedichev2003}
Fedichev, P.O.; Fischer, U.R.
Gibbons--Hawking effect in the sonic de Sitter space-time.
\textit{Phys.\ Rev.\ Lett.}\ \textbf{2003}, \textit{91}, 240407.

\bibitem{Fischer2004}
Fischer, U.R.; Sch\"utzhold, R.
Quantum simulation of cosmic inflation in two-component BECs.
\textit{Phys.\ Rev.\ A}\ \textbf{2004}, \textit{70}, 063615.

\bibitem{Barcelo2003}
Barcel\'o, C.; Liberati, S.; Visser, M.
Probing semiclassical analogue gravity in BECs.
\textit{Phys.\ Rev.\ A}\ \textbf{2003}, \textit{68}, 053613.

\bibitem{Horstmann2010}
Horstmann, B.; Reznik, B.; Fagnocchi, S.; Cirac, J.I.
Hawking radiation from an acoustic black hole on an ion ring.
\textit{Phys.\ Rev.\ Lett.}\ \textbf{2010}, \textit{104}, 250403.

\bibitem{Alsing2005}
Alsing, P.M.; Dowling, J.P.; Milburn, G.J.
Ion trap simulations of quantum fields in an expanding universe.
\textit{Phys.\ Rev.\ Lett.}\ \textbf{2005}, \mbox{\textit{94}, 220401.}

\bibitem{Schutzhold2007}
Schützhold, R.; Uhlmann, M.; Petersen, L.; Schmitz, H.; Friedenauer, A.; Schätz, T.
Analogue of cosmological particle creation in an ion trap.
\textit{Phys.\ Rev.\ Lett.}\ \textbf{2007}, \emph{99}, 201301.

\bibitem{Wittemer2019}
Wittemer, M.; Hakelberg, F.; Kiefer, P.; Schröder, J.P.; Fey, C.; Schützhold, R.; Warring, U.; Schaetz, T.
Particle pair creation by inflation of quantum vacuum fluctuations
in an ion trap.
\textit{Phys.\ Rev.\ Lett.}\ \textbf{2019}, \mbox{\emph{123}, 180502.}

\bibitem{Tian2017}
Tian, Z.; Jing, J.; Dragan, A.
Analog cosmological particle generation in a superconducting circuit.
\textit{Phys.\ Rev.\ D}\ \textbf{2017}, \emph{95}, 125003.

\bibitem{Lang2019}
Lang, S.; Sch\"utzhold, R.
Analog of cosmological particle creation in electromagnetic waveguides.
\textit{Phys.\ Rev.\ A}\ \textbf{2019}, \emph{100}, 052508.

\bibitem{Urtubey2026}
Urtubey, E.; Lombardo, F.C.; Villar, P.I.
Dissipative effects in transmission line analogues of Hawking radiation.
\textit{arXiv} \textbf{2026}, arXiv:2606.28505.

\bibitem{Nation2009}
Nation, P.D.; Blencowe, M.P.; Rimberg, A.J.; Buks, E.
Analogue Hawking radiation in a dc-SQUID array transmission line.
\textit{Phys.\ Rev.\ Lett.}\ \textbf{2009}, \emph{103}, 087004.

\bibitem{Schutzhold2005}
Sch\"utzhold, R.; Unruh, W.G.
Hawking radiation in an electromagnetic waveguide?
\textit{Phys.\ Rev.\ Lett.}\ \textbf{2005}, \emph{95}, 031301.

\bibitem{Eckel2018}
Eckel, S.; Kumar, A.; Jacobson, T.; Spielman, I.B.; Campbell, G.K.
A rapidly expanding Bose--Einstein condensate: An expanding universe in the lab.
\textit{Phys.\ Rev.\ X}\ \textbf{2018}, \emph{8}, 021021.

\bibitem{Viermann2022}
Viermann, C.; Sparn, M.; Liebster, N.; Hans, M.; Kath, E.; Parra-López, Á; Tolosa-Simeón, M.; Sánchez-Kuntz, N.; Haas, T.; Strobel, H.; et al.
Quantum field simulator for dynamics in curved spacetime.
\textit{Nature}\ \textbf{2022}, \emph{611}, 260.

\bibitem{Lombardo2012}
Lombardo, F.C.; Turiaci, G.J.
Decoherence and loss of entanglement in acoustic black holes.
\textit{Phys.\ Rev.\ Lett.}\ \textbf{2012}, \emph{108}, 261301.

\bibitem{Lombardo2013}
Lombardo, F.C.; Turiaci, G.J.
Quantum discord and decoherence in the cosmological horizon.
\textit{Phys.\ Rev.\ D}\ \textbf{2013}, \emph{87}, 084028.

\bibitem{Johansson2009}
Johansson, J.R.; Johansson, G.; Wilson, C.M.; Nori, F.
Dynamical Casimir effect in a superconducting coplanar waveguide.
\textit{Phys.\ Rev.\ Lett.}\ \textbf{2009}, \emph{103}, 147003.

\bibitem{Wilson2011}
Wilson, C.M.; Johansson, G.; Pourkabirian, A.; Simoen, M.; Johansson, J.R.; Duty, T.; Nori, F.; Delsing, P.
Observation of the dynamical Casimir effect in a superconducting circuit.
\textit{Nature}\ \textbf{2011}, \emph{479}, 376.

\bibitem{Lahteenmaki2013}
Lähteenmäki, P.; Paraoanu, G.S.; Hassel, J.; Hakonen, P.J.
Dynamical Casimir effect in a Josephson metamaterial.
\textit{Proc.\ Natl.\ Acad.\ Sci.\ USA}\ \textbf{2013}, \emph{110}, 4234.

\bibitem{Jaskula2012}
Jaskula, J.C.; Partridge, G.B.; Bonneau, M.; Lopes, R.; Ruaudel, J.; Boiron, D.; Westbrook, C.I.
Acoustic analog to the dynamical Casimir effect in a Bose--Einstein condensate.
\textit{Phys.\ Rev.\ Lett.}\ \textbf{2012}, \emph{109}, 220401.

\bibitem{Felicetti2014}
Felicetti, S.; Sanz, M.; Lamata, L.; Romero, G.; Johansson, G.; Delsing, P.; Solano, E.
Dynamical Casimir effect entangles artificial atoms.
\textit{Phys.\ Rev.\ Lett.}\ \textbf{2014}, \emph{113}, 093602.

\bibitem{Schwinger1951}
Schwinger, J. On gauge invariance and vacuum polarization.
\textit{Phys.\ Rev.}\ \textbf{1951}, \emph{82}, 664.

\bibitem{Rajeev2019}
Rajeev, K.; Chakraborty, S.; Padmanabhan, T.
Generalized Schwinger effect and particle production in an
expanding universe.
\textit{Phys.\ Rev.\ D}\ \textbf{2019}, \emph{100}, 045019.

\bibitem{Campo2005}
Campo, D.; Parentani, R.
Inflationary spectra and partially decohered distributions.
\textit{Phys.\ Rev.\ D}\ \textbf{2005}, \emph{72}, 045015.

\bibitem{Busch2014}
Busch, X.; Parentani, R.; Robertson, S.
Quantum entanglement due to a modulated dynamical Casimir effect.
\textit{Phys.\ Rev.\ A}\ \textbf{2014}, \mbox{\emph{89}, 063606}.

\bibitem{FranchinoVinas2026}
Franchino-Vi\~nas, S.A.; Mazzitelli, F.D.; Pla, S.
Conformally-flat gravitational analogues to the Schwinger effect.
\textit{arXiv} \textbf{2026}, arXiv:2602.18578.

\bibitem{Breuer2002}
Breuer, H.P.; Petruccione, F.
\textit{The Theory of Open Quantum Systems};
Oxford University Press: Oxford, UK, 2002.

\bibitem{CastellanosBeltran2008}
Castellanos-Beltran, M.A.; Irwin, K.D.; Hilton, G.C.; Vale, L.R.; Lehnert, K.W.
Amplification and squeezing of quantum noise with a tunable Josephson metamaterial.
\textit{Nat.\ Phys.}\ \textbf{2008}, \emph{4}, 929.

\bibitem{Braunstein2005}
Braunstein, S.L.; van Loock, P.
Quantum information with continuous variables.
\textit{Rev.\ Mod.\ Phys.}\ \textbf{2005}, \emph{77}, 513.

\bibitem{Polarski1996}
Polarski, D.; Starobinsky, A.A.
Semiclassicality and decoherence of cosmological perturbations.
\textit{Class.\ Quantum Grav.}\ \textbf{1996}, \emph{13}, 377.

\bibitem{Martin2016}
Martin, J.; Vennin, V.
Quantum discord of cosmic inflation: Can we show that CMB anisotropies are of quantum-mechanical origin?
\textit{Phys.\ Rev.\ D}\ \textbf{2016}, 93, 023505.

\bibitem{Krantz2019}
Krantz, P.; Kjaergaard, M.; Yan, F.; Orlando, T.P.; Gustavsson, S.; Oliver, W.D.
A quantum engineer's guide to superconducting qubits.
\textit{Appl.\ Phys.\ Rev.}\ \textbf{2019}, \emph{6}, 021318.

\bibitem{Albash2012}
Albash, T.; Boixo, S.; Lidar, D.A.; Zanardi, P.
Quantum adiabatic Markovian master equations.
\textit{New J.\ Phys.}\ \textbf{2012}, \textit{14}, 123016.
\end{thebibliography}
\end{document}